\documentclass[sigconf]{acmart}

\AtBeginDocument{%
  }

\usepackage{listings}
\usepackage{algorithm}
\usepackage[noend]{algpseudocode}  
\usepackage{dsfont}
\usepackage{xcolor}
\usepackage{float}
\usepackage{subcaption}
\usepackage{balance}
\floatstyle{ruled}
\newfloat{listing}{tbp}{lop}
\floatname{listing}{Listing}
\definecolor{codebg}{rgb}{0.95,0.95,0.95}
\definecolor{coderule}{rgb}{0.8,0.8,0.8}
\definecolor{keyword}{rgb}{0.0,0.0,0.6}
\definecolor{comment}{rgb}{0.25,0.5,0.35}
\definecolor{string}{rgb}{0.6,0.0,0.0}
\usepackage{enumitem}
\iftrue
\definecolor{darkgreen}{rgb}{0.0, 0.7, 0.0}
\newcommand{\izzat}[1]{{\color{darkgreen}[\textbf{\sc Izzat}: \textit{#1}]}}

\newcommand{\izzatrm}[1]{{\color{darkgreen}\sout{#1}}}
\else
\newcommand{\izzat}[1]{}

\newcommand{\izzatrm}[1]{}
\fi

\newcommand{\Th}{\boldsymbol{\theta}}

\theoremstyle{remark}

\theoremstyle{remark}

\theoremstyle{remark}

\usepackage{booktabs}
\usepackage{array}
\usepackage{xcolor}
\usepackage{colortbl}
\usepackage{graphicx}
\begin{document}

\title{\textsc{PruneX}: A Hierarchical Communication-Efficient System for Distributed CNN Training with Structured Pruning}

\author{Alireza Olama}
\email{alireza.olama@abo.fi}
\affiliation{%
  \institution{\AA bo Akademi University}
  \country{Finland}
}
\author{Andreas Lundell}
\email{andreas.lundell@abo.fi}
\affiliation{%
  \institution{\AA bo Akademi University}
  \country{Finland}
}
\author{Izzat El Hajj}
\email{izzat.elhajj@aub.edu.lb}
\affiliation{%
  \institution{American University of Beirut}
  \country{Lebanon}
}
\author{Johan Lilius}
\email{johan.lilius@abo.fi}
\affiliation{%
  \institution{\AA bo Akademi University}
  \country{Finland}
}
\author{Jerker Bj\"orkqvist}
\email{jerker.bjorkqvist@abo.fi}
\affiliation{%
  \institution{\AA bo Akademi University}
  \country{Finland}
}

\begin{abstract}
Inter-node communication bandwidth increasingly constrains distributed training at scale on multi-node GPU clusters. While compact models are the ultimate deployment target, conventional pruning-aware distributed training systems typically fail to reduce communication overhead because unstructured sparsity cannot be efficiently exploited by highly-optimized dense collective primitives. We present \textsc{PruneX}, a distributed data parallel training system that co-designs pruning algorithms with cluster hierarchy to reduce inter-node bandwidth usage. \textsc{PruneX} introduces Hierarchical Structured ADMM (H-SADMM) algorithm, which enforces node-level structured sparsity before inter-node synchronization, enabling dynamic buffer compaction that eliminates both zero-valued transmissions and indexing overhead. The system adopts a leader--follower execution model with separated intra-node and inter-node process groups, performing dense collectives on compacted tensors over bandwidth-limited links while confining full synchronization to high-bandwidth intra-node interconnects. Evaluation on ResNet architectures across 64 GPUs demonstrates that \textsc{PruneX} reduces inter-node communication volume by approximately 60\% and achieves 6.75$\times$ strong scaling speedup, outperforming the dense baseline (5.81$\times$) and Top-K gradient compression (3.71$\times$), on the Puhti supercomputer at CSC -- IT Center for Science (Finland).
\end{abstract}


\maketitle

\section{Introduction}
\label{sec:introduction}


The rapid growth of deep learning has been fueled by ever-larger datasets and increasingly complex models which necessitate distributed training systems with massive parallelism across Graphics Processing Unit (GPU) clusters~\cite{liang2024communication,amini2025distributed,narayanan2021efficient,
shoeybi2019megatron,rajbhandari2021zero}.
While frameworks like PyTorch DDP~\cite{li2020pytorch} and Horovod~\cite{sergeev2018horovod} maximize throughput via dense, synchronous aggregation, they consume enormous amounts of compute and communication bandwidth to train these models~\cite{liang2024communication,daily2018gossipgrad,
yu2023communication}.
At the same time, many deployments run on resource-constrained edge devices, where executing billions of floating-point operations is simply too costly~\cite{tu2025distributed,somvanshi2025tiny}. This has driven a strong need for structured sparsity that can be efficiently mapped to hardware accelerators and exploited through highly optimized dense kernels~\cite{cheng2024survey,he2023structured}.

There is a fundamental tension in current workflows because training is optimized for speed on large clusters while model compression is often left to a separate, post-hoc ‘train-prune-fine-tune’ stage~\cite{cheng2024survey,he2023structured}.
This approach is inherently inefficient since it consumes costly cluster resources on parameters that will ultimately be discarded. Moreover, because massive datasets are stored across distributed cloud systems, enabling pruning during single-node training is impractical. To overcome these challenges, model pruning must be integrated directly into the distributed training process, allowing efficiency from the start rather than as an afterthought.
\paragraph*{Related works and research gap}
Recent research has sought to address these challenges by introducing sparsity directly into the distributed training loop, generally following two main strategies.
The first strategy focuses on mitigating communication bottlenecks by sparsifying gradients before model updates. Methods such as Top-K and Deep Gradient Compression (DGC)~\cite{lin2018deep} achieve high theoretical sparsity ratios (up to 99.9\%) but rely on unstructured pruning. As shown in Table~\ref{tab:comparison}, methods relying on unstructured sparsity (Top-K, DGC) theoretically reduce data volume but often fail to accelerate training in practice~\cite{mishra2021accelerating,wang2020sparsert}. Recent studies demonstrate that gradient compression provides speedup in less than 3\% of data center scenarios due to the computational overhead of encoding and the reliance on non-scalable primitives like AllGather~\cite{agarwal2022utility}. Newer frameworks like Parallax~\cite{kim2019parallax} and EmbRace~\cite{li2022embrace} optimize communication for sparse natural language processing embeddings using hybrid architectures, yet they still incur indexing costs and fail to accelerate dense compute kernels on edge hardware due to irregular memory access patterns. PacTrain~\cite{wang2025pactrain} improves integration with \texttt{AllReduce} via mask tracking but remains bound by unstructured sparsity, offering limited computational speedup on standard accelerators.

As the second strategy, pruning-aware training algorithms attempt to reduce computational cost by learning sparse structures during optimization.
Approaches like PruneTrain~\cite{lym2019prunetrain} enforce structured sparsity (e.g., channel pruning) to enable efficient dense execution but rely on expensive `stop-and-reconfigure' mechanisms that periodically disrupt the training pipeline.
ClickTrain~\cite{zhang2021clicktrain} adopts fine-grained pattern pruning to balance flexibility and efficiency, yet its performance depends heavily on specialized compiler optimizations.
Moreover,  all existing frameworks treat the compute cluster as a flat topology, ignoring the bandwidth disparity between intra-node and inter-node links. 
\begin{table*}[t]
\caption{Comparison of \textsc{PruneX} against state-of-the-art distributed training frameworks.}
\centering
\label{tab:comparison}
\resizebox{\textwidth}{!}{%
\begin{tabular}{@{}lcccccc@{}}
\toprule
\textbf{Method} 
& \textbf{Sparsity Type} 
& \textbf{Metadata Overhead} 
& \textbf{Comm. Primitive} 
& \textbf{\shortstack{Topology\\Aware}} 
& \textbf{\shortstack{Communication Speedup\\(Network Efficiency)}} 
& \textbf{\shortstack{Inference Acceleration\\(GPU Compute Efficiency)}} \\ 
\midrule
Top-K / DGC~\cite{lin2018deep}
& Unstructured 
& High (Indices) 
& AllGather / Gossip 
& No 
& Low (Latency Bound) 
& None (Irregular Memory) \\

OmniReduce~\cite{fei2021efficient} 
& Unstructured 
& Medium (Bitmaps) 
& Streaming Agg. 
& No 
& Medium (Protocol Bound) 
& None (Sparse Indexing) \\

Parallax~\cite{kim2019parallax}  
& Unstructured 
& Low (Indices) 
& Hybrid (PS + AR) 
& No 
& High (Bandwidth) 
& None (Sparse Ops) \\

EmbRace~\cite{li2022embrace}  
& Unstructured 
& Medium (Indices) 
& Hybrid (AllToAll) 
& No 
& High (Overlap) 
& None (Sparse Ops) \\

PacTrain~\cite{wang2025pactrain}  
& Unstructured 
& Low (Mask Tracker) 
& \texttt{AllReduce}
& No 
& High (Dense Packed) 
& Low (Masked Ops) \\

ClickTrain~\cite{zhang2021clicktrain}  
& Fine-Grained 
& Low (Pattern Mask) 
& \texttt{AllReduce}
& No 
& High (Compiler Bound) 
& Medium (Compiler Dep.) \\

PruneTrain~\cite{lym2019prunetrain}  
& Structured 
& Zero (Reconfiguration) 
& \texttt{AllReduce}
& No 
& Medium (Reconfig Cost) 
& High (Dense Model) \\

\rowcolor{gray!20}
\textsc{PruneX}  
& Structured 
& Zero (Implicit) 
& Hierarchical AR 
& Yes 
& High (Shrunk Tensors) 
& High (Dense Kernels) \\

\bottomrule
\end{tabular}%
}
\end{table*}

A promising mathematical framework for structured pruning is the Alternating Direction Method of Multipliers (ADMM), which offers a rigorous method for enforcing geometric constraints such as filter or channel sparsity, primarily in centralized, single-device settings~\cite{boyd2011distributed,zhang2021structadmm,yuan2024reweighted,wang2024admm,lee2023compression}. 
This optimization-based approach offers theoretical advantages such as data-driven mask adaptation, and dynamic sparsity over the heuristic scheduling mechanisms employed by frameworks like PruneTrain and ClickTrain~\cite{zhang2021structadmm}.
However, standard distributed ADMM formulations are ill-suited for large-scale problems~\cite{olama2019relaxed,yang2022survey} and when applied to sparse optimization, sparsity is typically enforced on the global consensus variable \textit{after} node aggregation. Therefore, workers are forced to synchronize full-precision, dense parameter vectors across the network before the pruning step can be applied, effectively negating any potential communication benefits~\cite{olama2024distributed,olama2023tracking}.  To make structured pruning viable at scale, the optimization algorithm cannot be treated in isolation; it must be co-designed with the underlying physical topology to enforce sparsity before communication occurs.

This paper introduces \textsc{PruneX}, a hierarchical distributed training system that couples structured pruning with a consensus-ADMM formulation aligned to the cluster topology. \textsc{PruneX} implements a novel Hierarchical Structured ADMM (H-SADMM) algorithm that enforces structured sparsity on node-level consensus tensors. This enables a physical shrinkage mechanism in which node leaders compress active parameters into contiguous dense buffers before inter-node synchronization, achieving bandwidth savings without the metadata overhead of sparse formats. 

\paragraph*{Contributions} The main contributions of this paper are summarized as follows:
\begin{enumerate}
    \item We propose H-SADMM, a hierarchical consensus algorithm that enforces structured sparsity at the node level before inter-node communication, aligning optimization structure with cluster topology.
    
    \item We design a physical buffer shrinkage mechanism that compacts tensors using sparsity masks, enabling optimized dense collectives on reduced payloads without sparse-format metadata overhead.
    
    \item We implement a sparsity-aware leader-follower distributed architecture with separate intra-node and inter-node process groups, confining dense synchronization traffic to high-bandwidth local interconnects
    
    \item We release \textsc{PruneX} as open-source: \url{https://github.com/Alirezalm/PruneX}

\end{enumerate}

\paragraph{Paper Organization.} The remainder of this paper is organized as follows. In Section~\ref{sec:problem_formulation}, we formally describe the distributed training problem serving as the foundation for deriving the H-SADMM algorithm in Section~\ref{sec:hsadmm}. Section~\ref{sec:system_design} details the system architecture of \textsc{PruneX}. Finally, Section~\ref{sec:evaluation} presents our experimental evaluation, and Section~\ref{sec:conclusion} concludes the paper.

\begin{figure}
    \centering
    \begin{subfigure}[t]{0.48\linewidth}
        \centering
        \includegraphics[width=\linewidth]{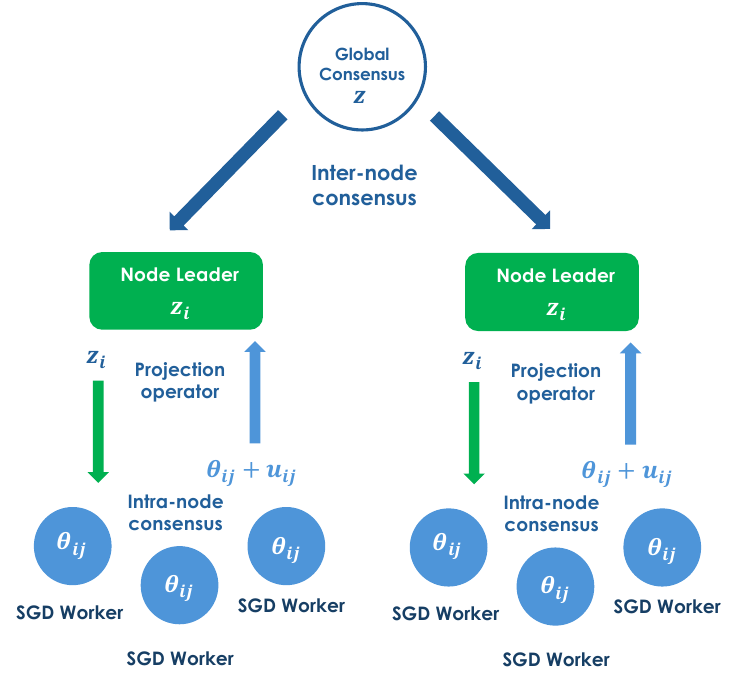}
        \caption{Hierarchical structure}
        \label{fig:hierarchical}
    \end{subfigure}\hfill
    \begin{subfigure}[t]{0.40\linewidth}
        \centering
        \includegraphics[width=\linewidth]{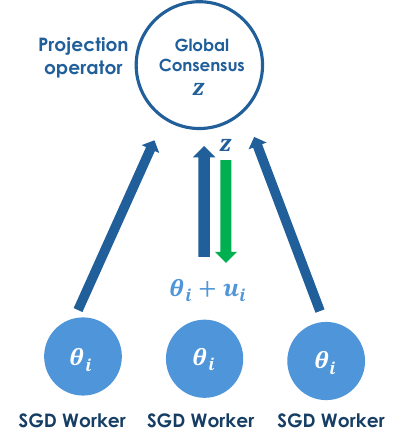}
        \caption{Flat structure}
        \label{fig:flat}
    \end{subfigure}
    \caption{A schematic comparison between (a) the proposed hierarchical consensus structure utilizing node leaders to bridge local and global states, and (b) the standard flat consensus structure where all workers communicate directly with the global variable.}
    \label{fig:structures}
\end{figure}
\section{Pruning-Aware Distributed Training Setup}
\label{sec:problem_formulation}
Consider a convolutional neural network (CNN) parameterized by a set of $L$ learnable weight tensors $\mathcal{W} = \{\mathbf{W}^\ell\}_{\ell=1}^{L}$ and corresponding bias vectors $\mathcal{B} = \{\mathbf{b}^\ell\}_{\ell=1}^{L}$. For a convolutional layer $\ell$, the weight tensor is defined as $\mathbf{W}^\ell \in \mathbb{R}^{C^\ell_{\text{out}} \times C^\ell_{\text{in}} \times K^\ell_H \times K^\ell_W}$, where $C^\ell_{\text{out}}$ and $C^\ell_{\text{in}}$ denote the output and input channel dimensions, and $(K^\ell_H, K^\ell_W)$ define the kernel's spatial height and width. For fully connected layers, the weights are defined as $\mathbf{W}^\ell \in \mathbb{R}^{d^\ell_{\text{out}} \times d^\ell_{\text{in}}}$.
%
%
Given input batch $\mathbf{X}$ and target labels $\mathbf{Y}$, the parameters $\Th = \{\mathcal{W}, \mathcal{B}\}$ are optimized by minimizing the empirical risk over the training dataset $\mathcal{D}$ by minimizing the loss function
\begin{equation}
f(\boldsymbol{\theta}) = \frac{1}{|\mathcal{D}|} \sum_{(\mathbf{X}, \mathbf{Y}) \in \mathcal{D}} \ell_{\text{ce}}\big(\Phi(\mathbf{X}; \boldsymbol{\theta}), \mathbf{Y}\big) + \frac{\lambda}{2} \sum_{\ell=1}^{L} \|\mathbf{W}^\ell\|_F^2,
\label{eq:loss_fn}
\end{equation}
where $\Phi(\mathbf{X}; \Th)$ defines model's forward mapping, $\lambda$ is the weight decay coefficient and $\ell_{\text{ce}}$ denotes the cross-entropy loss function. 




We adopt a data-parallel training paradigm on a hierarchical computing system consisting of $M$ compute nodes, indexed by $i \in \{1, \dots, M\}$. Each node is equipped with $P$ accelerators (\textit{e.g.}, GPUs), indexed by $j \in \{1, \dots, P\}$. 
The training dataset $\mathcal{D}$ is divided into $N$ disjoint, equally sized shards, with each shard stored exclusively on the $j$-th accelerator of the $i$-th node.
%
Consequently, the global objective function defined in Eq.~\eqref{eq:loss_fn} can be reformulated as the average of local objective functions as follows: 
%
%
%
%
\begin{equation}
\min_{\boldsymbol{\theta}} f(\boldsymbol{\theta}) = \frac{1}{N} \sum_{i=1}^{M} \sum_{j=1}^{P} f_{i,j}(\boldsymbol{\theta}) + \frac{\lambda}{2} \sum_{\ell=1}^{L} \|\mathbf{W}^\ell\|_F^2.
\label{eq:distributed_objective}
\end{equation}
where  $N = M \times P$ denotes the total number of accelerators.

\subsection{Structured Sparsity Constraints}\label{sec:sparsity-constraints}
To reduce computation and communication overhead, we impose structured sparsity on the convolutional layers $\mathcal{L}_{\text{conv}} \subset \{1,\dots,L\}$. Filter sparsity controls the number of active output channels by defining a vector $\mathbf{m}_f \in \mathbb{R}^{C^\ell_{\text{out}}}$ with entries $(\mathbf{m}_f)_k = \|\mathbf{W}_{\ell,k,:,:,:}\|_F$, and restricting the weight tensor to $\mathcal{S}_f^\ell = \{\mathbf{W}^\ell \mid \|\mathbf{m}_f\|_0 \le \alpha^{\ell}_{\text{out}}\}$.
Channel sparsity similarly limits the number of active input channels using $\mathbf{m}_c \in \mathbb{R}^{C^\ell_{\text{in}}}$ where $(\mathbf{m}_c)_k = \|\mathbf{W}_{\ell,:,k,:,:}\|_F$, resulting in $\mathcal{S}_c^\ell = \{\mathbf{W}^\ell \mid \|\mathbf{m}_c\|_0 \le \alpha^{\ell}_{\text{in}}\}$. Shape sparsity constrains the spatial receptive field by defining a magnitude tensor $\mathcal{M}_s \in \mathbb{R}^{C^\ell_{\text{in}} \times K^\ell_H \times K^\ell_W}$ with $(\mathcal{M}_s)_{b,u,v} = \|\mathbf{W}_{\ell,:,b,u,v}\|_F$ and restricting the weights to $\mathcal{S}_s^\ell = \{\mathbf{W}^\ell \mid \|\text{vec}(\mathcal{M}_s)\|_0 \le \alpha^{\ell}_{\text{kern}}\}$. 
%
We define the composite structured sparsity set $\mathcal{S}^\ell$ as the intersection of all applicable constraints as $\mathcal{S}^\ell = 
\mathcal{S}_f^\ell \cap \mathcal{S}_c^\ell \cap \mathcal{S}_s^\ell.$
%
%
For layers where a specific constraint is not enforced, the corresponding set is treated as the unconstrained parameter space.


\subsection{Constrained Distributed Optimization}
Combining the distributed objective formulation from Eq.~\eqref{eq:distributed_objective} with the sparsity constraints defined by $\mathcal{S}^\ell$, the final constrained optimization problem solved during training is formulated as
\begin{equation}
\begin{aligned}
& \min_{\boldsymbol{\theta}} \quad \frac{1}{N} \sum_{i=1}^{M} \sum_{j=1}^{P} f_{i,j}(\boldsymbol{\theta}) + \frac{\lambda}{2} \sum_{\ell=1}^{L} \|\mathbf{W}^\ell\|_F^2 \\
& \text{s.t.} \quad \mathbf{W}^\ell \in \mathcal{S}^\ell, \quad \ell = 1, \dots, L.
\end{aligned}
\label{eq:constrained_problem}
\end{equation}
This formulation seeks a global model that simultaneously minimizes the regularized empirical risk and satisfies the pruning constraints imposed by the sparsity sets $\mathcal{S}^\ell$.


\subsection{Hierarchical Consensus Reformulation}

To enable fully decoupled parallel optimization and exploit the hierarchical nature of modern data center hardware, we reformulate problem~\eqref{eq:constrained_problem} by introducing local copies of the model parameters. Let $\boldsymbol{\theta}_{i,j}$ denote the local model parameters residing on the $j$-th accelerator of the $i$-th node. In a standard distributed setting, all local copies are constrained to agree directly with a global variable \cite{boyd2011distributed,olama2024distributed} as illustrated in Figure~\ref{fig:flat}.
However, this flat consensus model fails to account for the bandwidth disparity between intra-node interconnects (\textit{e.g.}, NVLink) and inter-node networks (\textit{e.g.}, Ethernet or InfiniBand). To address this limitation, we decompose the global consensus into a two-level hierarchy involving intermediate node-level consensus variables $\mathbf{z}_i = \{\mathbf{z}_i^\ell\}_{\ell=1}^L$ and global consensus variables $\mathbf{z} = \{\mathbf{z}^\ell\}_{\ell=1}^L$. This introduces two distinct constraints: intra-node consensus, where $\boldsymbol{\theta}_{i,j}^\ell = \mathbf{z}_i^\ell$ for all accelerators $j$ within node $i$; and inter-node consensus, where $\mathbf{z}_i^\ell = \mathbf{z}^\ell$ across node leaders (see Figure~\ref{fig:hierarchical}).

Crucially, we enforce the structured sparsity constraints $\mathcal{S}^\ell$ explicitly on the node-level variables $\mathbf{z}_i^\ell$ rather than solely on the global variable. The reformulated problem is given by
\begin{equation}
\begin{aligned}
& \min_{\{\boldsymbol{\theta}_{i,j}\}, \{\mathbf{z}_i\}, \mathbf{z}} \quad \frac{1}{N} \sum_{i=1}^{M} \sum_{j=1}^{P} f_{i,j}(\boldsymbol{\theta}_{i,j}) + \frac{\lambda}{2} \sum_{\ell=1}^{L} \|\mathbf{z}^\ell\|_F^2 \\
& \text{s.t.} \quad \boldsymbol{\theta}_{i,j}^\ell = \mathbf{z}_i^\ell, \quad \forall i \in \{1,\dots,M\}, j \in \{1,\dots,P\}, \ell \in \{1,\dots,L\}, \\
& \phantom{\text{s.t.}} \quad \mathbf{z}_i^\ell = \mathbf{z}^\ell, \quad \forall i \in \{1,\dots,M\}, \ell \in \{1,\dots,L\}, \\
& \phantom{\text{s.t.}} \quad \mathbf{z}_i^\ell \in \mathcal{S}^\ell, \quad \forall i \in \{1,\dots,M\}, \ell \in \{1,\dots,L\}.
\end{aligned}
\label{eq:hierarchical_problem}
\end{equation}

This specific placement of the sparsity constraint provides an algorithmic advantage over standard pruning-aware distributed ADMM approaches \cite{olama2024distributed,olama2019relaxed} where the sparsity is typically enforced after model synchronization.
In contrast, our hierarchical formulation enforces $\mathbf{z}_i^\ell \in \mathcal{S}^\ell$ at the node level. This ensures that the node-level consensus variables are sparse \textit{before} inter-node synchronization. Since the constraints in $\mathcal{S}^\ell$ (e.g., filter or channel pruning) physically reduce tensor dimensions, we can utilize significantly smaller communication buffers for the inter-node exchange of $\mathbf{z}_i^\ell$.

\section{Hierarchical Structured ADMM (H-SADMM)}
\label{sec:hsadmm}
In this section, we introduce our proposed H-SADMM algorithm for solving Eq.~\eqref{eq:hierarchical_problem}. For any structured sparsity set $\mathcal{S}^\ell$, we use the indicator function $\mathbb{I}_{\mathcal{S}^\ell}(\mathbf{v})$, which equals $0$ if $\mathbf{v} \in \mathcal{S}^\ell$ and $+\infty$ otherwise~\cite{boyd2011distributed}. This reformulation allows us to embed the sparsity constraints directly into the objective function. Equivalently, problem~\eqref{eq:hierarchical_problem} can be written as
\begin{equation}
\begin{aligned}
& \min_{\{\boldsymbol{\theta}_{i,j}\}, \{\mathbf{z}_i\}, \mathbf{z}} \ \frac{1}{N} \sum_{i=1}^{M} \sum_{j=1}^{P} f_{i,j}(\boldsymbol{\theta}_{i,j}) + \sum_{i=1}^{M} \sum_{\ell=1}^{L} \left( \frac{\lambda}{2M} \|\mathbf{z}_i^\ell\|_F^2 + \mathbb{I}_{\mathcal{S}^\ell}(\mathbf{z}_i^\ell) \right) \\
& \text{s.t.} \quad \boldsymbol{\theta}_{i,j}^\ell = \mathbf{z}_i^\ell, \quad \forall i, j, \ell, \\
& \phantom{\text{s.t.}} \quad \mathbf{z}_i^\ell = \mathbf{z}^\ell, \quad \forall i, \ell.
\end{aligned}
\label{eq:reformulated_problem}
\end{equation}

%
To solve problem Eq.~\eqref{eq:reformulated_problem}, we construct the Augmented Lagrangian in the scaled form by introducing two sets of layer-wise penalty parameters, $\boldsymbol{\rho}_1 = \{\rho_1^\ell\}_{\ell=1}^L$ and $\boldsymbol{\rho}_2 = \{\rho_2^\ell\}_{\ell=1}^L$, corresponding to the intra-node and inter-node constraints, respectively.
Let $\mathbf{u}_{i,j}^\ell$ denote the scaled dual variables associated with the intra-node constraints ($\boldsymbol{\theta}_{i,j}^\ell = \mathbf{z}_i^\ell$), and let $\mathbf{v}_i^\ell$ denote the scaled dual variables associated with the inter-node constraints ($\mathbf{z}_i^\ell = \mathbf{z}^\ell$).
The Augmented Lagrangian function $L_{\boldsymbol{\rho}}(\{\boldsymbol{\theta}_{i,j}\}, \{\mathbf{z}_i\}, \mathbf{z}, \{\mathbf{u}_{i,j}\}, \{\mathbf{v}_i\})$ is defined as
\begin{equation}
\begin{aligned}
L_{\boldsymbol{\rho}} &= \frac{1}{N} \sum_{i=1}^{M} \sum_{j=1}^{P} f_{i,j}(\boldsymbol{\theta}_{i,j}) + \sum_{i=1}^{M} \sum_{\ell=1}^{L} \left( \frac{\lambda}{2M} \|\mathbf{z}_i^\ell\|_F^2 + \mathbb{I}_{\mathcal{S}^\ell}(\mathbf{z}_i^\ell) \right) \\
&\quad + \sum_{i=1}^{M} \sum_{j=1}^{P} \sum_{\ell=1}^{L} \frac{\rho_1^\ell}{2} \left\| \boldsymbol{\theta}_{i,j}^\ell - \mathbf{z}_i^\ell + \mathbf{u}_{i,j}^\ell \right\|_F^2 \\
&\quad + \sum_{i=1}^{M} \sum_{\ell=1}^{L} \frac{\rho_2^\ell}{2} \left\| \mathbf{z}_i^\ell - \mathbf{z}^\ell + \mathbf{v}_i^\ell \right\|_F^2,
\end{aligned}
\label{eq:augmented_lagrangian}
\end{equation}
where constant terms involving only the dual variables are omitted for brevity.

Our proposed H-SADMM algorithm solves this optimization problem by applying Block Coordinate Descent (BCD) to the Augmented Lagrangian \cite{boyd2011distributed,tseng2001convergence}. Instead of jointly optimizing all variables, the BCD iteratively minimizes $L_{\rho}$ with respect to one block of variables
while holding the others fixed. In the following subsections, we discuss each minimization step.

\subsection{Minimization Step for \texorpdfstring{$\boldsymbol{\theta}$}{Theta}}

The first step of the H-SADMM algorithm involves minimizing ~\eqref{eq:augmented_lagrangian} with respect to $\Th_{i,j}$, which decomposes into $N$ independent sub-problems that can be solved in parallel by each accelerator.
For a specific accelerator $j$ on node $i$ at iteration $k$, the update rule is given by
\begin{equation}
\boldsymbol{\theta}_{i,j}^{k+1} = \arg\min_{\boldsymbol{\theta}} \left( f_{i,j}(\boldsymbol{\theta}) + \sum_{\ell=1}^{L} \frac{\rho_1^\ell}{2} \left\| \boldsymbol{\theta}^\ell - \mathbf{z}_i^{\ell, k} + \mathbf{u}_{i,j}^{\ell, k} \right\|_F^2 \right).
\label{eq:theta_update}
\end{equation}

Problem \eqref{eq:theta_update} can be interpreted as standard local training with an additional proximal regularization term. The first term, $f_{i,j}(\boldsymbol{\theta})$, drives the parameters to minimize the empirical risk on the local dataset shard. The second term is a quadratic penalty that enforces consistency with the current node-level consensus variable $\mathbf{z}_i^k$, shifted by the accumulated dual residual $\mathbf{u}_{i,j}^k$.
%
We approximate the solution of Problem~\eqref{eq:theta_update} using multiple steps of SGD on the local dataset. Initializing $\Th_{i,j}^{k,0} = \boldsymbol{\theta}_{i,j}^k$, we perform $E$ local epochs, and for each mini-batch $\xi$ the inner update at iteration $\tau$ is
\begin{equation}
\boldsymbol{\theta}_{i,j}^{k, \tau+1} = \boldsymbol{\theta}_{i,j}^{k, \tau} - \eta \left( \nabla f_{i,j}(\boldsymbol{\theta}_{i,j}^{k, \tau}; \xi) + \boldsymbol{\rho}_1 \odot (\boldsymbol{\theta}_{i,j}^{k, \tau} - \mathbf{z}_i^k + \mathbf{u}_{i,j}^k) \right),
\label{eq:theta_sgd}
\end{equation}
where $\eta$ is the learning rate, $\nabla f_{i,j}(\cdot; \xi)$ is the stochastic gradient of the loss function computed on mini-batch $\xi$, and $\odot$ denotes element-wise multiplication broadcasting the layer-wise penalties $\boldsymbol{\rho}_1$ across the parameter tensors. After $E$ epochs, we set $\boldsymbol{\theta}_{i,j}^{k+1} = \boldsymbol{\theta}_{i,j}^{k, \tau_{\text{end}}}$.

\subsection{Minimization Step for $\mathbf{z}_i$}

The second step of the H-SADMM algorithm updates the node-level consensus variables $\mathbf{z}_i$. This step is critical as it enforces the structured sparsity constraints defined in $\mathcal{S}^\ell$ before any inter-node communication occurs.

Because of the quadratic structure of the $\ell_2$-regularized objective and the indicator function in~\eqref{eq:augmented_lagrangian}, the solution of this step can be decoupled into an unconstrained quadratic minimization over $\mathbf{z}_i$ followed by a Euclidean projection of $\mathbf{z}_i$ onto the sparsity set $\mathcal{S}^\ell$.
%
Let $\gamma^\ell = \frac{\lambda}{M} + P\rho_1^\ell + \rho_2^\ell$ be the normalization factor. The dense candidate solution is given by the weighted average of the local updates and the global consensus
\begin{equation}
\tilde{\mathbf{z}}_i^{\ell, k+1} = \frac{1}{\gamma^\ell} \left[ \rho_1^\ell \sum_{j=1}^{P} (\boldsymbol{\theta}_{i,j}^{\ell, k+1} + \mathbf{u}_{i,j}^{\ell, k}) + \rho_2^\ell (\mathbf{z}^{\ell, k} - \mathbf{v}_i^{\ell, k}) \right].
\label{eq:z_dense_candidate}
\end{equation}

Second, we obtain the sparse node-level variable by applying the projection operator $\Pi_{\mathcal{S}^\ell}(\cdot)$:
\begin{equation}
\mathbf{z}_i^{\ell, k+1} = \Pi_{\mathcal{S}^\ell} \big( \tilde{\mathbf{z}}_i^{\ell, k+1} \big).
\label{eq:z_projection}
\end{equation}

The projection operator $\Pi_{\mathcal{S}^\ell}(\mathbf{X})$ finds the tensor in $\mathcal{S}^\ell$ closest to $\mathbf{X}$ in terms of the Frobenius norm. We compute the closed-form solutions for these projections for each sparsity set, following the methodology in \cite{zhang2021structadmm}.

When multiple constraints are enforced simultaneously (e.g., $\mathcal{S}^\ell = \mathcal{S}_f^\ell \cap \mathcal{S}_c^\ell$), the projection $\Pi_{\mathcal{S}^\ell}$ is computed by applying the individual projections sequentially. Since the structural groups (rows vs. columns) are orthogonal in the GEMM representation, the order of projection does not affect the final support set.


\subsection{Minimization Step for Global \texorpdfstring{$\mathbf{z}$}{z}}

The third step of the H-SADMM algorithm updates the global consensus variable $\mathbf{z}$. This step enforces consistency across different compute nodes.
The optimization of the Augmented Lagrangian in Eq.~\eqref{eq:augmented_lagrangian} with respect to $\mathbf{z}$ reduces to minimizing an unconstrained convex quadratic, whose closed-form solution is given by
\begin{equation}
\mathbf{z}^{\ell, k+1} = \frac{1}{M} \sum_{i=1}^{M} \left( \mathbf{z}_i^{\ell, k+1} + \mathbf{v}_i^{\ell, k} \right).
\label{eq:global_z_update}
\end{equation}

\subsection{Dual Variable Updates}

The intra-node dual variables $\mathbf{u}_{i,j}$ capturing the mismatch between local accelerators and the node-level consensus, are updated as
\begin{equation}
\mathbf{u}_{i,j}^{\ell,k+1} = \mathbf{u}_{i,j}^{\ell,k} + \left( \boldsymbol{\theta}_{i,j}^{\ell,k+1} - \mathbf{z}_i^{\ell,k+1} \right).
\end{equation}
The inter-node dual variables $\mathbf{v}_i$ tracking discrepancies between the node-level and global models, follow
\begin{equation}
\mathbf{v}_i^{\ell,k+1} = \mathbf{v}_i^{\ell,k} + \left( \mathbf{z}_i^{\ell,k+1} - \mathbf{z}^{\ell,k+1} \right).
\end{equation}


We monitor convergence using layer-wise primal and dual residuals at both hierarchy levels. For intra-node consensus, the primal and dual residuals are $\mathbf{r}_{\text{intra},i,j}^{\ell,k+1}=\boldsymbol{\theta}_{i,j}^{\ell,k+1}-\mathbf{z}_i^{\ell,k+1}$ and $\mathbf{s}_{\text{intra},i,j}^{\ell,k+1}=\rho_1^\ell(\mathbf{z}_i^{\ell,k+1}-\mathbf{z}_i^{\ell,k})$, while inter-node consensus uses $\mathbf{r}_{\text{inter},i}^{\ell,k+1}=\mathbf{z}_i^{\ell,k+1}-\mathbf{z}^{\ell,k+1}$ and $\mathbf{s}_{\text{inter},i}^{\ell,k+1}=\rho_2^\ell(\mathbf{z}^{\ell,k+1}-\mathbf{z}^{\ell,k})$. 
Following standard ADMM practice~\cite{boyd2011distributed}, convergence for layer $\ell$ is declared once the total primal and dual residual norms fall below the feasibility thresholds $\epsilon_{\text{pri}}^\ell$ and $\epsilon_{\text{dual}}^\ell$, defined using the absolute and relative tolerances described in~\cite{boyd2011distributed}. To balance residual decay, we adopt a layer-wise variant of adaptive penalty tuning introduced in~\cite{boyd2011distributed}.  This update is applied independently to $\rho_1^\ell$ and $\rho_2^\ell$, allowing each layer’s penalties to adjust according to synchronization difficulty.

\label{subsec:mask_sync}

\section{System Design and Implementation}
\label{sec:system_design}

In this section, we describe the architecture and implementation of \textsc{PruneX}, guided by three design goals: 
(i) Hierarchy-aware orchestration, where intra-node and inter-node consensus levels align with the physical network hierarchy.
(ii) Sparsity-enabled bandwidth reduction, in which sparsity masks serve as system primitives that determine communication buffer sizes, ensuring inter-node bandwidth scales with model density rather than total parameter count. 
(iii) Dense-kernel compatibility, achieved by leveraging structured sparsity and avoiding unstructured sparse formats with indexing overhead and relying on optimized dense collectives (e.g., NCCL \texttt{AllReduce}~\cite{hu2025demystifying,nccl2019}).
%

    \begin{figure}
    \centering
    \includegraphics[width=\linewidth]{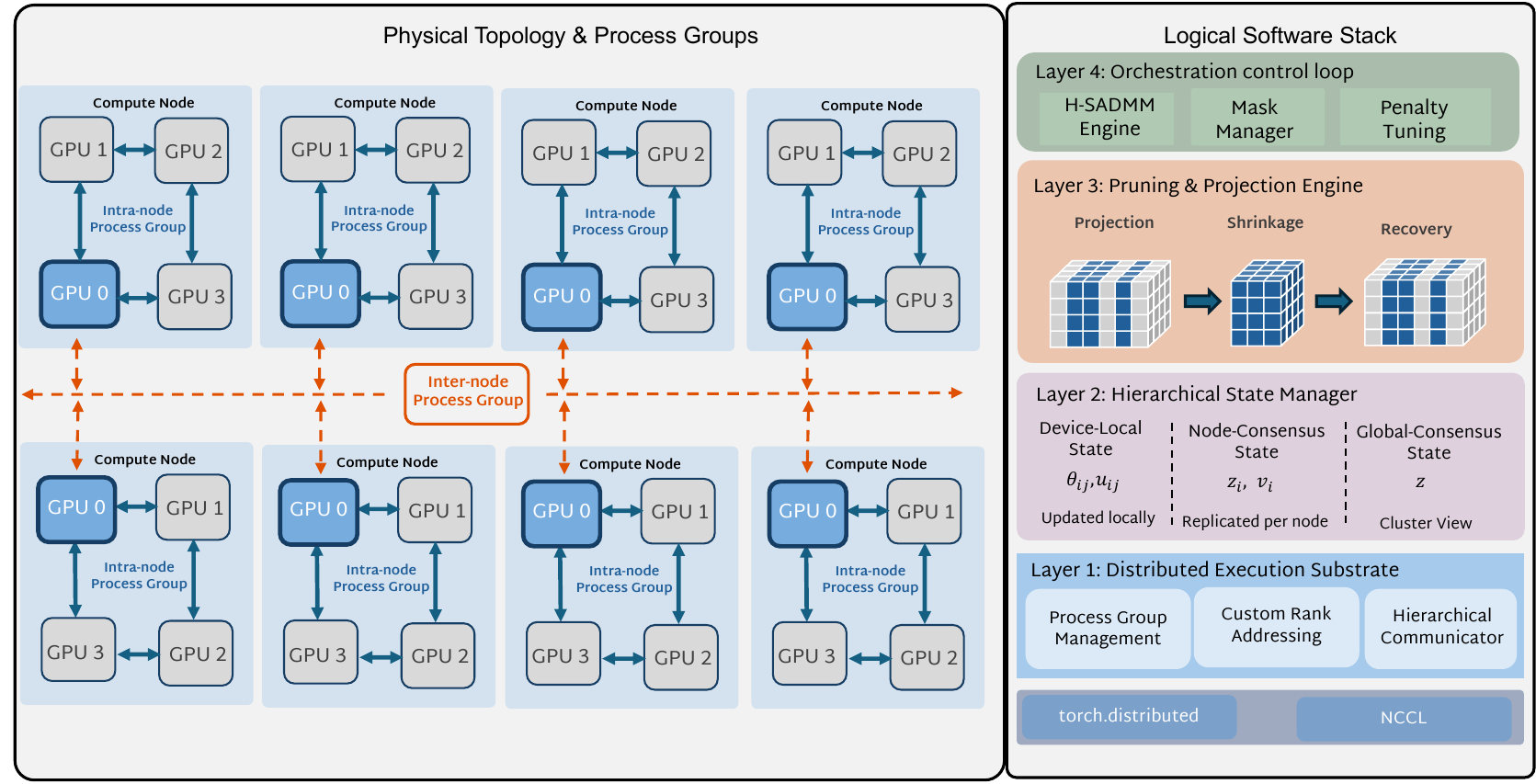}
    \caption{\textsc{PruneX} System Architecture. The system design maps the physical cluster topology to a four-layer logical software stack, comprising the Distributed Execution Substrate, Hierarchical State Manager, Pruning Engine, and Orchestration Control Loop.}
    \label{fig:architecture}
\end{figure}

\subsection{General Architecture}
\label{subsec:architecture}

\textsc{PruneX} is architected as a layered system over a standard deep learning execution substrate (\textit{e.g.}, PyTorch Distributed). We consider a homogeneous cluster environment where each node $M$ is equipped with an identical number of accelerators $P$. This symmetry allows for a static mapping of global ranks to physical resources, simplifying the orchestration of hierarchical collectives. The system is organized into four primary layers as depicted in Figure~\ref{fig:architecture}.

\subsubsection{Distributed Execution Substrate}
\label{sec:distributed-substrate}
To address the bandwidth disparity between intra-node interconnects (e.g., NVLink) and inter-node networks (e.g., InfiniBand), \textsc{PruneX} adopts a hierarchical \textit{leader-follower} \texttt{AllReduce} strategy that aligns the execution flow with the physical cluster topology. Unlike standard flat strategies (e.g., Ring-AllReduce) that treat all processes symmetrically, our approach decouples local synchronization from global consensus. 
%
Upon initialization, the system performs topology discovery to map global ranks to physical hardware resources. We implement a custom process group management layer over the standard execution backend (e.g., \texttt{torch.distributed} with NCCL). The system automatically assigns roles based on locality. \textit{Workers} are processes responsible for local SGD updates on individual accelerators. \textit{Node Leaders} are the processes with local rank zero on each node, which manage the interface between the node's consensus state and the rest of the cluster (see Figure~\ref{fig:architecture}).
To strictly enforce this separation of concerns, \textsc{PruneX} instantiates two distinct levels of communicators using the \texttt{new\_group} API. \textit{Intra-node Groups} connect all workers within a single node to rapidly synchronize the node-level consensus variables $z_i$. \textit{Inter-node Groups} connect only the node leaders across the cluster-wide fabric. By restricting global membership to leaders, we reduce the number of participants in the inter-node \texttt{AllReduce} phase.
As it will be explained in Section~\ref{subsec:system_mapping}, the node leader in \textsc{PruneX} is not merely a passive router but an active \textit{pruning gateway}. In standard hierarchical implementations, leaders simply aggregate and forward dense tensors~\cite{jia2018highly}. In \textsc{PruneX}, the leader serves as the integration point for the H-SADMM constraints which enforce the projection operator $\Pi_{S_l}$ and apply the "physical shrinkage" mechanism (see Section \ref{subsec:shrinkage}). This ensures that the data entering the Inter-node Group is already compressed into dense, contiguous buffers of active parameters.
\subsubsection{Hierarchical State Manager}
Layered above the communication substrate, the State Manager maps the mathematical variables of the H-SADMM algorithm directly onto the physical process hierarchy. This component partitions the optimization state space into three distinct tiers with strict storage and update ownership rules.
First, the \textit{Device-Local State}—comprising model parameters $\boldsymbol{\theta}_{i,j}$ and intra-node dual variables $\mathbf{u}_{i,j}$—resides in high-bandwidth device memory. These tensors are allocated independently on each worker and are updated strictly via parallel local compute kernels, requiring no communication. Second, the \textit{Node-Consensus State} ($\mathbf{z}_i, \mathbf{v}_i$) represents a logically shared node-level view. While these variables are physically replicated across all workers within node $i$ to facilitate parallel downstream computation, their consistency is maintained via high-frequency intra-node \texttt{AllReduce} operations. Finally, the \textit{Global-Consensus State} ($\mathbf{z}$) represents the cluster-wide view managed by the Inter-node Group. Node leaders update $\mathbf{z}$ via sparsity-aware inter-node aggregation and subsequently propagate the result to followers via intra-node \texttt{Broadcast}s. This strict ownership model establishes a clear data movement contract: only the node-level variables $\mathbf{z}_i$ and $\mathbf{v}_i$ are candidates for inter-node transmission, and only after being projected onto the low-dimensional sparsity manifolds.

\subsubsection{Pruning and Projection Engine}
This component serves as a middleware layer between the optimization logic and the communication subsystem. Its role is to enforce the structured constraints $\mathcal{S}^\ell$ while intercepting the node-consensus state $\mathbf{z}_i$ to perform projection (pruning) and to generate the corresponding sparsity masks. In addition, the engine reconfigures inter-node communication buffers by packing only the active tensor slices into dense, contiguous memory blocks before transmission through the network stack. Detailed descriptions of the mask synchronization and shrinkage procedures are provided in Section~\ref{subsec:mask_sync} and Section~\ref{subsec:shrinkage}, respectively.

\subsubsection{Orchestration Control Loop}
The top-level controller manages the execution flow of H-SADMM, coordinating the bottom-up propagation of updates (Local Training $\to$ Node Aggregation $\to$ Mask Generation $\to$ Global Aggregation) and the top-down \texttt{Broadcast} of global consensus. Detailed descriptions of the system mapping are provided in Section~\ref{subsec:system_mapping}. This layered architecture explicitly separates compute-intensive local training from latency-sensitive consensus protocols, enabling efficient scaling by leveraging the bandwidth hierarchy of the underlying physical cluster.

%
\subsubsection{Extensibility to Deeper Hierarchies}
While the current implementation targets a two-tier hierarchy (node and cluster), the process model is designed to generalize to deeper hierarchies topologies, such as rack-scale aggregations. In such configurations, the system can introduce intermediate consensus variables (e.g., $\mathbf{z}_{\text{rack}}$) and corresponding process groups. The modular separation between process group management and the optimization state ensures that such extensions require no fundamental redesign of the core pruning or communication logic.
    \begin{figure}
    \centering
    \includegraphics[width=1\linewidth]{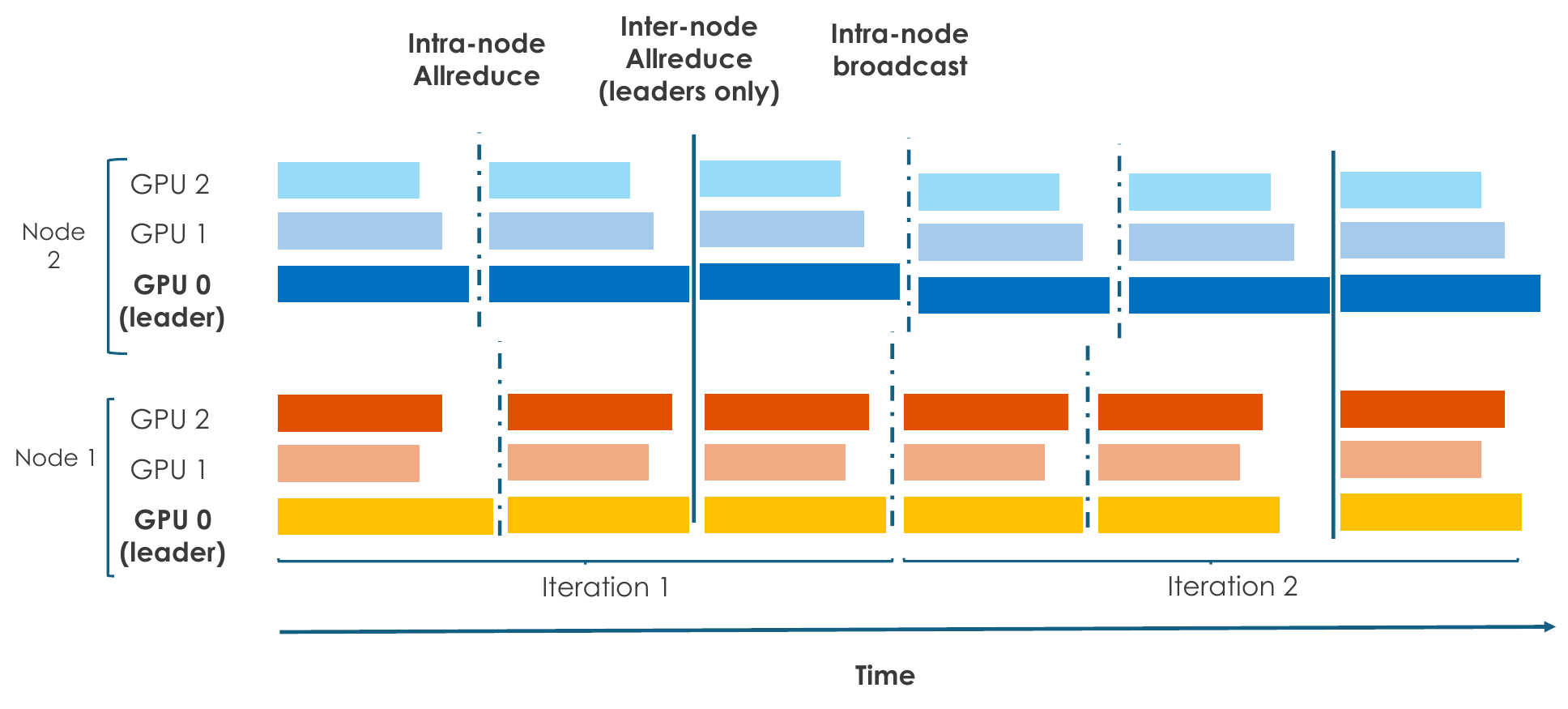}
    \caption{Runtime Execution Timeline. The schedule illustrates the overlapping of high-bandwidth intra-node synchronization (\texttt{AllReduce} and \texttt{Broadcast}) with local computation, while latency-sensitive inter-node communication is isolated to the node leaders.}
    \label{fig:time_line}
\end{figure}
\subsection{Runtime Execution Flow}
\label{subsec:system_mapping}

The \textsc{PruneX} runtime serializes the mathematical dependencies of the H-SADMM algorithm into an execution sequence consisting of five distinct phases detailed in Algorithm~\ref{alg:prunex_execution}. Figure~\ref{fig:time_line} provides a visual timeline of this execution flow.
\begin{algorithm}[t]
\caption{\textsc{PruneX} H-SADMM Algorithm}
\label{alg:prunex_execution}
\begin{algorithmic}[1]
\scriptsize
\Require Dataset shards $\mathcal{D}_{i,j}$, Sparsity Constraints $\mathcal{S}$, Process Groups $\mathcal{G}_{\text{intra}}, \mathcal{G}_{\text{inter}}$
\Require Hyperparameters: $\rho_1, \rho_2$, Freeze Epoch $T_{\text{freeze}}$

\State \textbf{Initialize:} $\boldsymbol{\theta}_{i,j}, \mathbf{z}_i, \mathbf{z} \leftarrow \text{Random/Pre-trained}$;\quad $\mathbf{u}_{i,j}, \mathbf{v}_i \leftarrow \mathbf{0}$; \quad $\mathbf{m}_{\text{global}} \leftarrow \mathbf{1}$

\For{iteration $k = 1$ \textbf{to} $K_{\text{max}}$}
    \State \textcolor{darkgray}{// Phase 1: Local Computation (Parallel on all Accelerators)}
    \State $\boldsymbol{\theta}_{i,j}^{k+1} \leftarrow \text{SGD}(\nabla f_{i,j}, \boldsymbol{\theta}_{i,j}^k, \mathbf{z}_i^k, \mathbf{u}_{i,j}^k)$ \label{line:sgd}\Comment{Local Training}
    
    \State \textcolor{darkgray}{// Phase 2: Intra-node Consensus (High-Bandwidth) ---}
    \State $\mathbf{buf} \leftarrow \boldsymbol{\theta}_{i,j}^{k+1} + \mathbf{u}_{i,j}^k$ \label{line:buffer}
    \State $\bar{\mathbf{z}}_i \leftarrow \textsc{AllReduce}(\mathbf{buf}, \textsc{Sum}, \mathcal{G}_{\text{intra}})$ \label{line:intra-all-reduce}\Comment{Compute dense node candidate}
    \State $\tilde{\mathbf{z}}_i \leftarrow \textsc{UpdateIntra}(\bar{\mathbf{z}}_i,\mathbf{z},\mathbf{v}_i)$ \Comment{Eq. \eqref{eq:z_dense_candidate}}
    
    \State \textcolor{darkgray}{// Phase 3: Node Projection \& Mask Generation}
    \If{$k > T_{\text{freeze}}$} 
        \State $\mathbf{z}_i^{k+1} \leftarrow \tilde{\mathbf{z}}_i \odot \mathbf{m}_{\text{global}}$ \Comment{\textbf{Retraining:} Apply frozen static mask}
    \Else
        \State $\mathbf{z}_i^{k+1} \leftarrow \Pi_{\mathcal{S}}(\tilde{\mathbf{z}}_i)$ \label{line:projection} \Comment{\textbf{Dynamic:} Project onto sparsity set}
        \State $\mathbf{m}_{i} \leftarrow \mathcal{M}(\mathbf{z}_i^{k+1})$ \label{line:mask-gen}
    \EndIf

    \State \textcolor{darkgray}{// Phase 4: Inter-node Consensus (Leader-Only, Low-Bandwidth)}
    \If{\textbf{IsLeader}$(rank)$}
        
            \If{$k \le T_{\text{freeze}}$}
                \State $\mathbf{m}_{\text{global}} \leftarrow \textsc{AllReduce}(\mathbf{m}_{i}, \textsc{BitwiseOr}, \mathcal{G}_{\text{inter}})$ \Comment{Sync masks} \label{line:mask_sync}
            \EndIf
            
            \State $\mathbf{c}_{\text{dense}} \leftarrow \mathbf{z}_i^{k+1} + \mathbf{v}_i^k$
            \State $\mathbf{c}_{\text{shrunk}} \leftarrow \text{Compress}(\mathbf{c}_{\text{dense}}, \mathbf{m}_{\text{global}})$ \Comment{Physically remove pruned params} \label{line:shrunk}
            \State $\hat{\mathbf{z}} \leftarrow \textsc{AllReduce}(\mathbf{c}_{\text{shrunk}}, \textsc{Avg}, \mathcal{G}_{\text{inter}})$ \Comment{Dense collective on small buffer}\label{line:inter-all-reduce}
            \State $\mathbf{z}^{k+1} \leftarrow \text{Decompress}(\hat{\mathbf{z}}, \mathbf{m}_{\text{global}})$ \Comment{Restore tensor shape with zeros}\label{line:decompress}
            
            \State $\mathbf{v}_i^{k+1} \leftarrow \mathbf{v}_i^k + (\mathbf{z}_i^{k+1} - \mathbf{z}^{k+1})$ \Comment{Update inter-node dual variables}
            \State $t_{\text{sync}} \leftarrow k$
        
        \State \textbf{Broadcast} $\mathbf{z}^{k+1}$ to $\mathcal{G}_{\text{intra}}$ \label{line:bcast}\Comment{Propagate global view to workers}
    \EndIf

    \State \textcolor{darkgray}{// Phase 5: Dual Update \& Convergence Check}
    \State $\mathbf{u}_{i,j}^{k+1} \leftarrow \mathbf{u}_{i,j}^k + (\boldsymbol{\theta}_{i,j}^{k+1} - \mathbf{z}_i^{k+1})$ \Comment{Update intra-node dual variables}
    
    \State $R_{\text{pri}}, R_{\text{dual}} \leftarrow \text{ComputeGlobalResiduals}(\dots)$
    \If{$R_{\text{pri}} < \epsilon^{\text{pri}}$ \textbf{and} $R_{\text{dual}} < \epsilon^{\text{dual}}$}
        \State \textbf{Break} \Comment{Convergence achieved}
    \EndIf
    \State Update $\rho_1, \rho_2$ \Comment{Adaptive penalty tuning}
\EndFor
\end{algorithmic}
\end{algorithm}
The iteration starts with the \textit{Local Computation Phase}, which corresponds to the minimization of $\boldsymbol{\theta}_{i,j}$. In this stage, each worker independently executes a fixed number of SGD epochs on its local data shard (line~\ref{line:sgd}). The standard loss function is augmented on-the-fly with the quadratic penalty term $\frac{\rho_1^\ell}{2} \| \boldsymbol{\theta}_{i,j}^\ell - \mathbf{z}_i^{\ell, k} + \mathbf{u}_{i,j}^{\ell, k} \|_F^2$. This phase utilizes the full computational throughput of the accelerators and requires no communication.
Upon completion of local updates, the system transitions to the \textit{Intra-node Consensus Phase}. This phase resolves the dense node-level candidate $\tilde{\mathbf{z}}_i$ required for the projection step. Each worker constructs a temporary buffer (line~\ref{line:buffer}) combining its local parameters and intra-node duals, $\boldsymbol{\theta}_{i,j} + \mathbf{u}_{i,j}$, and participates in an \texttt{AllReduce} sum operation over the high-bandwidth Intra-node Group (line~\ref{line:intra-all-reduce}). Because this operation occurs over fast interconnects, it is latency-bound rather than bandwidth-bound, incurring minimal synchronization overhead.

The third stage is the \textit{Node-level Projection and Mask Synchronization Phase}. Unlike standard data-parallel training which would immediately aggregate dense gradients globally, \textsc{PruneX} enforces a \textit{projection barrier} (line~\ref{line:projection}). Each worker uses the aggregated dense candidate to compute the updated node-consensus variable $\mathbf{z}_i$ by applying the projection operator $\Pi_{\mathcal{S}^\ell}$. This derives the binary sparsity masks that define the active structural support (line~\ref{line:mask-gen}). To ensure a consistent global topology for the subsequent aggregation, these local masks are synchronized across the cluster via bit-wise operations (line~\ref{line:mask_sync}); the detailed procedure for this mask synchronization step is described in Section~\ref{subsec:mask_sync}. The outcome is a unified global mask for each layer, agreed upon by all leaders, which dictates the shape of the tensors to be exchanged.

The fourth phase, \textit{Inter-node Consensus and Hierarchical Communication}, addresses the system's primary bandwidth bottleneck. Node leaders construct the inter-node buffers by combining the projected node-consensus variables with the inter-node duals, $\mathbf{z}_i + \mathbf{v}_i$. Crucially, before entering the network stack, leaders apply the global masks to physically compact these tensors, discarding pruned elements (line~\ref{line:shrunk}); the detailed procedure for this tensor shrinkage step is described in Section~\ref{subsec:shrinkage}. The leaders then execute the global \texttt{AllReduce} over the Inter-node Group using these shrunk buffers (line~\ref{line:inter-all-reduce}). The resulting compact global consensus $\mathbf{z}$ is then broadcast top-down from leaders to workers within each node, where it is scattered back into the full tensor shape for the next iteration (lines~\ref{line:decompress} and \ref{line:bcast}).
The iteration concludes with the \textit{Dual Update and Residual Computation Phase}. This memory-bound stage updates the state variables to prepare for the next cycle. Device-level dual variables $\mathbf{u}_{i,j}$ are updated based on the divergence between local parameters and the node consensus, while node-level dual variables $\mathbf{v}_i$ track the drift between the node and global consensus. Concurrently, the system computes the distributed primal and dual residual norms to drive the adaptive penalty tuning logic.

\subsection{Bitwise Mask Synchronization}
\label{subsec:mask_sync}


Following the node-level projection phase, individual nodes may arrive at divergent sparsity patterns due to variations in local data distributions and learning dynamics. To enable the subsequent dense inter-node aggregation, the system must reconcile these local views into a single, consistent global topology. This is achieved through a specialized synchronization protocol that generates the global binary mask $\mathbf{m}^\ell$ from the set of local node masks $\{\mathbf{m}_i^\ell\}_{i=1}^M$.

The synchronization proceeds in two steps designed to be robust and bandwidth-efficient. First, the boolean mask tensors are derived from the projection operator. Second, the system invokes a global \texttt{AllReduce} with a logical \texttt{OR} operator across the cluster. This operation effectively computes the logical union of all local masks: 
\begin{equation}
\mathbf{m}^\ell = \bigvee_{i=1}^{M} \mathbf{m}_i^\ell.
\end{equation}
Finally, this unified mask is stored in the hierarchical state manager to guide the subsequent communication and optimization steps.


\begin{figure}[htbp]
    \centering
    \includegraphics[width=\linewidth]{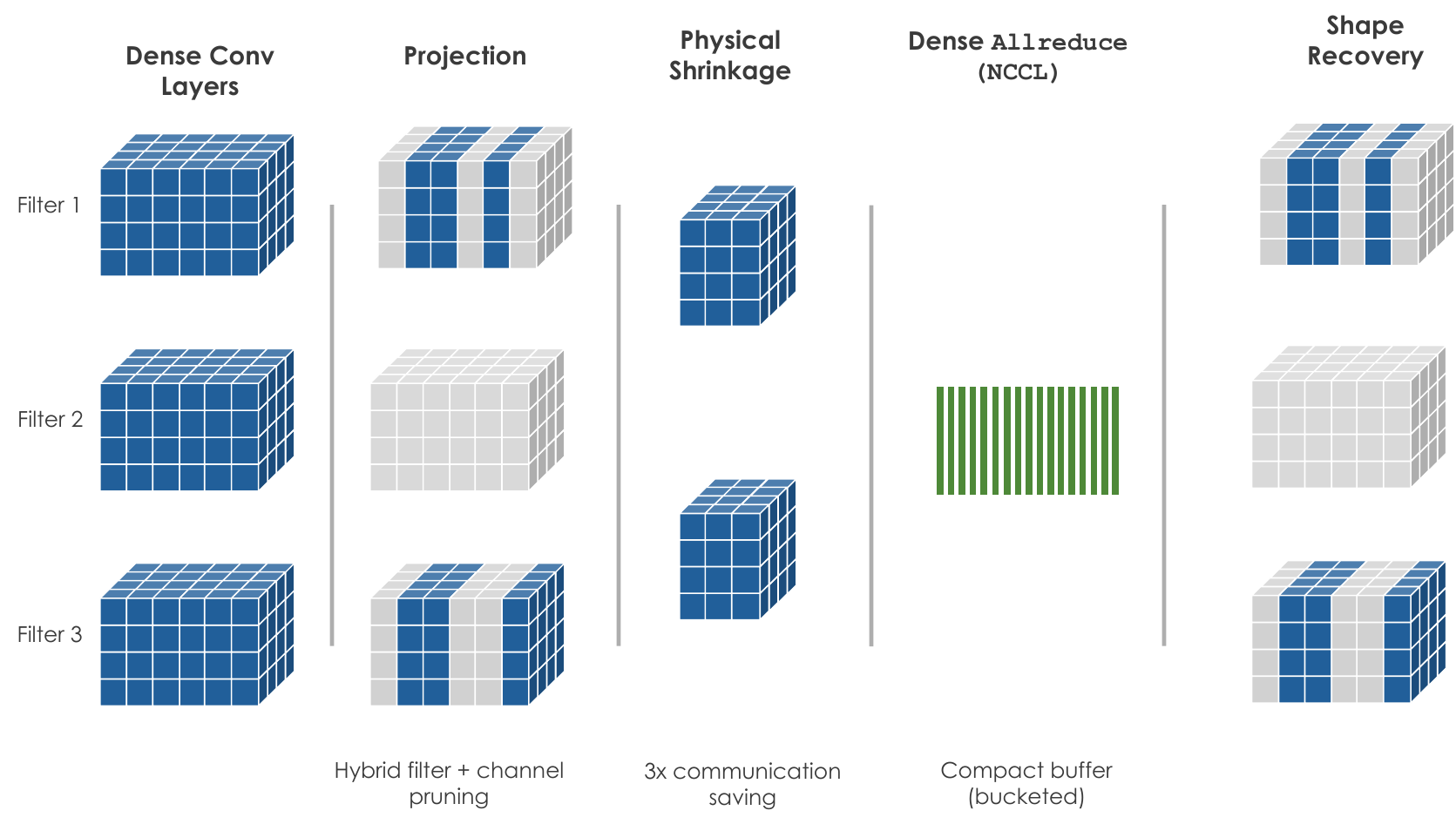}
    \caption{Physical Shrinkage and Recovery Pipeline. The mechanism transforms sparse tensors into compact dense buffers using global masks, executes the inter-node \texttt{AllReduce} on reduced payloads, and restores the full tensor shape via zero-filling for subsequent local training.}
    \label{fig:shrinkage}
\end{figure}
\subsection{Physical Shrinkage and Recovery Mechanism}
\label{subsec:shrinkage}

The key advantage of \textsc{PruneX} for satisfying goal (ii) (Sparsity-Enabled Bandwidth Reduction) is the physical shrinkage of inter-node communication buffers. While standard sparse communication methods often rely on transmitting indices alongside values (e.g., COO format), which incurs significant metadata overhead and processing latency, our approach leverages the globally synchronized masks to restructure tensors into dense, contiguous blocks prior to transmission. The complete pipeline for this physical shrinkage and subsequent shape recovery is illustrated in Figure~\ref{fig:shrinkage}.

\subsubsection{Tensor Shrinkage}
Shrinkage is carried out at the node leaders immediately before the inter-node consensus step. For each layer $\ell$, the leader first forms the pre-transmission buffer by combining the projected node-level consensus tensor and the inter-node dual variable, i.e., $\mathbf{c}_i^\ell = \mathbf{z}_i^\ell + \mathbf{v}_i^\ell$. For convolutional layers, $\mathbf{c}_i^\ell$ is a four-dimensional tensor of shape $(C_{\text{out}}^\ell \times C_{\text{in}}^\ell \times K_H \times K_W)$, and the system uses the globally synchronized mask $\mathbf{m}^\ell$ to determine which structured slices remain active and are therefore worth communicating. Concretely, the implementation derives two one-dimensional index sets from the mask: an output-filter set $\mathcal{K}_{\text{out}}^\ell$ containing those filters for which at least one masked entry is active, and an input-channel set $\mathcal{K}_{\text{in}}^\ell$ containing those channels for which at least one masked entry is active. In code, these sets are obtained by boolean reductions (\texttt{any}) over the spatial dimensions (and the complementary channel/filter dimension) followed by extraction of the nonzero entries; these are standard PyTorch tensor operations and run on the GPU whenever the mask and buffers reside on CUDA.
Once $\mathcal{K}_{\text{out}}^\ell$ and $\mathcal{K}_{\text{in}}^\ell$ are computed, the leader physically packs a dense compact tensor by slicing the original 4D buffer along the kept output and input dimensions, producing
\begin{equation}
\hat{\mathbf{c}}_i^\ell \;=\; \mathbf{c}_i^\ell\big[\mathcal{K}_{\text{out}}^\ell,\mathcal{K}_{\text{in}}^\ell,:,:\big],
\end{equation}
with shape $(|\mathcal{K}_{\text{out}}^\ell| \times |\mathcal{K}_{\text{in}}^\ell| \times K_H \times K_W)$. The implementation realizes this packing using PyTorch advanced indexing and then materializes the result into contiguous memory (via a \texttt{.contiguous()} call). This packing step is inherently memory-bandwidth bound because it gathers and copies selected structured slices into a new compact buffer; however, it is still beneficial because it produces a contiguous dense tensor that can be communicated using optimized dense collectives.
Dense collectives require shape agreement across all participants, and the system enforces this by ensuring that $\mathbf{m}^\ell$ is synchronized globally before shrinkage as described in Section~\ref{subsec:shrinkage}. As a result, all leaders compute identical $\mathcal{K}_{\text{out}}^\ell$ and $\mathcal{K}_{\text{in}}^\ell$ and therefore produce compacted tensors of identical shapes for each layer.

\subsubsection{Compacted Global Aggregation}
When an inter-node synchronization is scheduled, the leaders execute the inter-node consensus directly on the compact buffers produced by the shrinkage operator. In the implementation, this step is realized using a standard dense \texttt{AllReduce} over the inter-node leader group followed by a division by the number of participating leaders.
To improve the communication efficiency, we implement a \textit{bucketed} \texttt{AllReduce} strategy inspired by PyTorch Distributed's gradient bucketing design~\cite{li2020pytorch}. Specifically, compact payloads are coalesced into contiguous communication buffers of approximately 32MB before the collective operation is invoked, as illustrated by the equally spaced rectangles in Figure~\ref{fig:shrinkage}. This synchronization strategy amortizes per-call latency and kernel-launch overheads over larger messages, thereby increasing the effective bandwidth utilization of the inter-node fabric and reducing tail latency effects caused by numerous small layer-wise collectives~\cite{li2020pytorch}. 
\subsubsection{Recovery and Consistency}
After the inter-node \texttt{AllReduce} completes on the compact buffers, each leader obtains the compact consensus tensor $\hat{\mathbf{z}}^\ell$ for each convolutional layer. To restore the original tensor layout required by subsequent local computation, the system expands $\hat{\mathbf{z}}^\ell$ back to the full convolution shape using the cached index mappings $\mathcal{K}_{\text{out}}^\ell$ and $\mathcal{K}_{\text{in}}^\ell$ computed during shrinkage. The recovery operator allocates a zero-initialized tensor $\mathbf{z}^\ell$ of the original size $(C_{\text{out}}^\ell \times C_{\text{in}}^\ell \times K_H \times K_W)$ and scatters the compact values into the coordinates determined by the Cartesian product of the kept output and input indices, leaving all other entries exactly zero. 
Equivalently, for any $(o,c,h,w)$,
\begin{equation*}
\mathbf{z}^\ell[o,c,h,w] \;=\;
\begin{cases}
\hat{\mathbf{z}}^\ell[p,q,h,w] & \text{if } o=\mathcal{K}_{\text{out}}^\ell[p]\ \text{and}\ c=\mathcal{K}_{\text{in}}^\ell[q], \\
0 & \text{otherwise}.
\end{cases}
\end{equation*}
In the implementation, this zero-fill expansion is performed without explicit Python loops by using a vectorized indexed scatter (PyTorch’s \texttt{index\_put\_}) into the zero-filled destination tensor, and it executes on the same device as the tensors (GPU if available).
Because the mask is globally synchronized, all leaders and followers share identical index sets, so the recovered full-sized tensors are shape-consistent and semantically identical across nodes.
%


\subsection{Mask Freezing and Retraining Phase}
\label{subsec:retraining}

To ensure convergence to a high-accuracy solution, \textsc{PruneX} transitions from dynamic structure discovery to static fine-tuning. As the H-SADMM penalty parameters increase and optimization proceeds, the sparsity patterns naturally stabilize; empirically, mask drift converges to zero within the first 5--15 iterations (see Figure~\ref{fig:comm_volume}). Leveraging this behavior, the system activates the \textit{Mask Freezing Protocol} once topological stability is detected or when a predefined epoch $T_{\text{freeze}}$ is reached. At this stage, the global masks $\{\mathbf{m}^\ell\}$ are fixed, effectively locking the model's structural support for the remainder of training.
The projection operator $\Pi_{\mathcal{S}^\ell}$ is replaced by a low-cost element-wise application of the cached static masks. Consequently, the shapes of the compacted inter-node tensors $\hat{\mathbf{c}}_i^\ell$ become invariant, enabling \textit{One-Shot Buffer Allocation} where communication buffers are pre-allocated once and reused. This eliminates the overhead of dynamic shape inference and memory re-allocation, reducing the optimization problem to standard distributed training on a fixed sparse manifold.



\section{Evaluation}
\label{sec:evaluation}
In this section, we present an empirical evaluation of \textsc{PruneX} using the Puhti supercomputer at CSC -- IT Center for Science (Finland)~\cite{puhti2025}. Our goal is to assess the practical effectiveness of the proposed system under realistic distributed training conditions. To this end, we analyze \textsc{PruneX} along multiple complementary dimensions, including end-to-end training throughput, communication efficiency across the network hierarchy, scalability with increasing numbers of GPUs, and the convergence behavior of the underlying optimization algorithm. Together, these experiments are designed to provide a comprehensive view of both the system-level and algorithmic benefits of \textsc{PruneX} in large-scale distributed settings.

\subsection{Evaluation Methodology}
\label{subsec:setup}

\subsubsection{Software specifications}
\label{subsec:implementation}
We implemented \textsc{PruneX} using PyTorch (v2.9.1) as the primary deep learning framework~\cite{paszke2019pytorch}. The distributed control plane is built on top of the \texttt{torch.distributed} package (included in PyTorch 2.9.1), leveraging the NVIDIA NCCL backend (v2.25.1) for high-performance collective communication on CUDA-enabled GPUs (CUDA 12.8). To support the hierarchical process groups described in Section~\ref{subsec:architecture}, we utilize PyTorch's \texttt{new\_group} API to instantiate separate communicators for intra-node and inter-node traffic.

\subsubsection{Hardware specifications}
We evaluate \textsc{PruneX} using the Puhti supercomputer, specifically its Puhti-AI partition comprising 80 GPU nodes. Each node is equipped with four NVIDIA Volta V100 GPUs~\cite{choquette2020nvidia} (32,GB memory each), connected via NVLink intra-node interconnects. The nodes are interconnected via a low-latency Mellanox HDR InfiniBand fabric, providing 100~Gbps bandwidth per link. Each node is powered by two Intel Xeon Gold 6230 (Cascade Lake) CPUs with 20 cores per socket.
The experiments are conducted on a configuration of 16 nodes, utilizing a total of 64 GPUs ($16$ nodes $\times$ $4$ GPUs/node), unless otherwise specified.
\subsubsection{Datasets}
For our distributed training workloads, we use the CIFAR-10 dataset~\cite{krizhevsky2009learning} and train three deep residual network architectures: ResNet-152, ResNet-18~\cite{he2016deep}, and WideResNet-50-2~\cite{zagoruyko2016wide}  (see Table \ref{tab:model_specs}). ResNet-152 serves as our primary benchmark for scalability and communication analysis due to its significant depth and parameter count.
\begin{table}[h]
\centering
\caption{Evaluated Model Architectures.}
\label{tab:model_specs}
\resizebox{0.9\linewidth}{!}{%
\begin{tabular}{@{}l c c c@{}}
\toprule
\textbf{CNN Architecture} & \textbf{Number of Layers} & \textbf{Number of Parameters} & \textbf{GFLOPs} \\
\midrule
\textbf{ResNet-18} & 18 & $\sim$11 M & 1.8 \\
\textbf{WideResNet-50-2} & 50 & $\sim$69 M & 11.4 \\
\textbf{ResNet-152} & 152 & $\sim$60 M & 11.3 \\
\bottomrule
\end{tabular}%
}
\end{table}



\subsubsection{Baselines}\label{sec:baselines}
To evaluate the performance of \textsc{PruneX}, we compare it against two distinct baselines representing the state-of-the-art in distributed training:
\begin{itemize}[leftmargin=*]
    \item \textit{PyTorch DDP (Dense Baseline)}: We use the standard Pytorch \texttt{DistributedDataParallel} (DDP) module as the primary baseline for throughput and accuracy~\cite{li2020pytorch}. DDP performs dense synchronous SGD, aggregating full-precision gradients across all workers at every iteration using a ring-\texttt{AllReduce} algorithm.
    \item \textit{Top-K SGD (Gradient Compression Baseline)}: To benchmark against prior work, we implement a distributed Top-$K$ gradient compression baseline~\cite{lin2018deep}. We configure the sparsification rate to $0.01$, ensuring that each worker transmits strictly the top $1\%$ of gradient entries by magnitude per layer, while maintaining all other hyperparameters consistent with the dense DDP baseline.
    \item \textit{PruneX (AR):} To isolate the performance gains attributable to our system architecture, we evaluate an ablation of \textsc{PruneX} that implements the flat consensus structure depicted in~\ref{fig:flat}. In this configuration, all workers communicate directly with the global variable, bypassing the leader-follower hierarchy. Consequently, this baseline does not benefit from the sparsity-enabled buffer shrinkage or the hierarchical bandwidth optimization, serving as a direct reference to quantify the utility of our hierarchy-aware design
\end{itemize}

\subsubsection{Hyperparameter Tuning and Reproducibility}
\label{subsec:tuning}
We report the key hyperparameters and tuning choices used to perform the experiments. Unless stated otherwise, our experiments use a batch size of $128$ per GPU and SGD with learning rate $10^{-3}$, momentum $0.9$ and weight decay $10^{-4}$. For H-SADMM, we perform up to 60 outer iterations depending on convergence, with 5–10 inner (local) training epochs per iteration, which is typically sufficient for convergence. We initialize the intra-node and inter-node penalty parameters to $\rho_1=1.5\times 10^{-3}$ and $\rho_2=1.5\times 10^{-4}$, and cap them at $\rho_1^{\max}=10.0$ and $\rho_2^{\max}=10.0$. Structured sparsity is applied only to convolutional layers: in the primary configuration we use channel pruning with keep rate $0.5$ (i.e., $50\%$ channel density), while the implementation also supports filter pruning as well as composing multiple sparsity operators (e.g., filter and channel jointly) by intersecting their masks.

\subsection{End-to-End Training Efficiency}
\label{subsec:eval_training_speed}

The ultimate metric for system performance is the \textit{time-to-accuracy} which is the wall-clock time required to train a model to a target validation accuracy. We evaluate this holistic metric in Figure~\ref{fig:time_to_acc}, comparing \textsc{PruneX} against the DDP and Top-K baselines on the ResNet-152 workload.
\subsubsection{Time-to-Accuracy}
\begin{figure}[t]
    \centering
    \includegraphics[width=\linewidth]{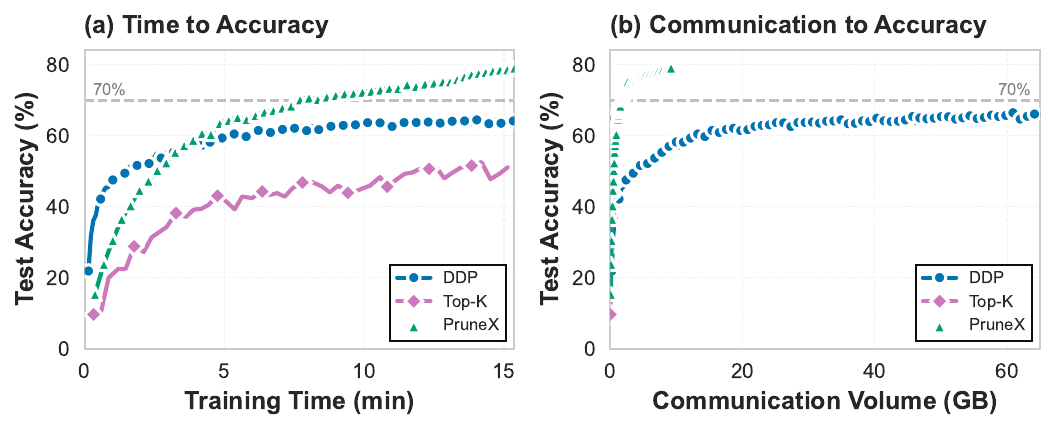}
    \caption{End-to-End Training Efficiency on ResNet-152. (a) \textsc{PruneX} (Green) reaches the 70\% target accuracy faster than DDP (Blue) and Top-K (Pink), validating that algorithmic overhead is outweighed by communication gains. (b) Accuracy vs. Inter-node Communication Volume demonstrates that \textsc{PruneX} achieves high accuracy with a fraction of the data transfer required by dense training.
    }
    \label{fig:time_to_acc}
\end{figure}

Figure~\ref{fig:time_to_acc}(a) plots the test accuracy against cumulative training time. The standard DDP baseline exhibits a robust initial learning curve but is constrained by the high communication overhead of synchronizing the dense 60 M parameters of ResNet-152, reaching only ${\sim}65\%$ accuracy within the 14-minute window. The Top-K baseline, while communicating less data, suffers from the noise induced by gradient sparsification and error accumulation, resulting in significantly slower convergence and lower final accuracy (${\sim}50\%$).

In contrast, \textsc{PruneX} demonstrates superior efficiency. Although it incurs a slight initial overhead due to the auxiliary variable initialization and the warm-up of the ADMM penalties, it rapidly surpasses the baselines. \textsc{PruneX} crosses the $70\%$ accuracy threshold in approximately 8 minutes, a target that the DDP baseline fails to reach within the allocated time budget. This result confirms that the computational cost of the local projection ($\Pi_{\mathcal{S}^\ell}$) and mask synchronization is negligible compared to the time saved by reducing inter-node data transfer.

\subsubsection{Sample Efficiency per Byte}
To isolate the bandwidth efficiency from computational throughput, Figure~\ref{fig:time_to_acc}(b) visualizes the test accuracy as a function of the inter-node cumulative communication volume (in GB).
The dense DDP baseline accumulates over 60 GB of traffic to reach its peak accuracy. \textsc{PruneX}, however, exhibits a strictly steeper ascent, reaching the $70\%$ accuracy target with significantly less cumulative data transfer. This metric effectively represents the `accuracy per byte' efficiency of the system, proving that the structured sparsity filters out redundant parameters before they consume network resources.

\subsection{Communication Efficiency Analysis}
\label{subsec:eval_communication}
We evaluate the ability of \textsc{PruneX} to reduce network utilization and hide communication latency during training, which directly addresses design goals (i) and (ii).

\subsubsection{Bandwidth Reduction via Physical Shrinkage}

\begin{figure}[t]
    \centering
    \includegraphics[width=\linewidth]{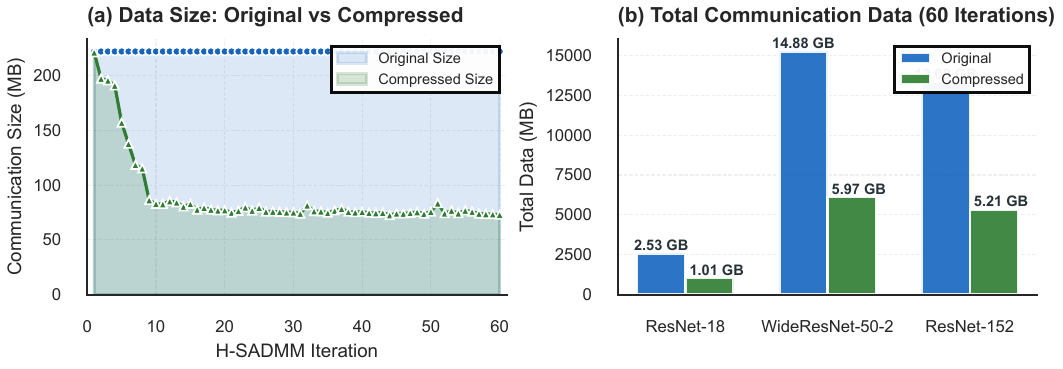}
    \caption{Data Volume Analysis on ResNet-152, ResNet18, and WideResNet-50-2 on 64 GPUs. (a) The compressed message size per H-SADMM iteration (ResNet-152). (b) Total communication volume reduction.}
    \label{fig:comm_volume}
\end{figure}

Figure~\ref{fig:comm_volume} quantifies the impact of the physical shrinkage mechanism on the total data volume exchanged over the inter-node fabric. Figure~\ref{fig:comm_volume}(a) tracks the message size per iteration for ResNet-152. Initially, the compressed size mirrors the original dense size (${\sim}220$ MB) as the sparsity masks are initialized to all-ones. As the H-SADMM penalties increase and the projection operator $\Pi_{\mathcal{S}^\ell}$ begins pruning filters, the compressed payload size drops precipitously, stabilizing at approximately $40\%$ of the original volume.
Figure~\ref{fig:comm_volume}(b) aggregates this reduction over the full training trajectory across three ResNet architectures. The system demonstrates consistent savings across model sizes: ResNet-18 communication volume drops from 2.53 GB to 1.01 GB, while the larger WideResNet-50-2 decreases from 14.88 GB to 5.97 GB. Similarly, for ResNet-152, \textsc{PruneX} reduces the total inter-node traffic from 13.00 GB (Dense DDP) to 5.21 GB. This corresponds to a consistent ${\sim}60\%$  reduction in communication volume across all evaluated models.

\subsubsection{Latency Hiding via Hierarchical Consensus}
\begin{figure}[t]
    \centering
    \includegraphics[width=\linewidth]{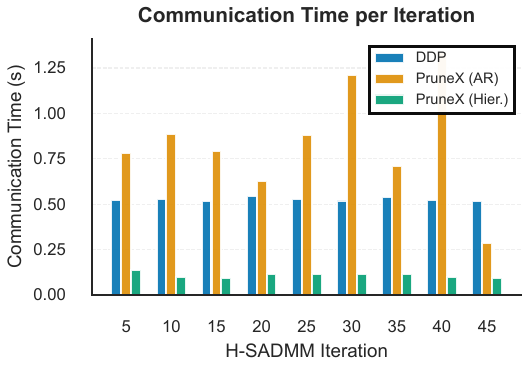}
    \caption{Per-Iteration Communication Latency. Comparison of communication latency per iteration, highlighting the speedup of the hierarchical \textsc{PruneX} configuration (blue) against DDP (red) and  flat PruneX \texttt{AllReduce} baseline (yellow).}
    \label{fig:comm_time}
\end{figure}

We next analyze how this volume reduction translates to wall-clock time. Figure~\ref{fig:comm_time} compares the per-iteration communication latency of \textsc{PruneX} against PyTorch DDP. The dense baseline exhibits a constant latency of approximately $0.50$s per iteration. In contrast, the hierarchical configuration of \textsc{PruneX} achieves a communication latency of roughly $0.10$s, representing a $5\times$ speedup in the synchronization phase. 
Crucially, the 'PruneX (AR)' baseline suffers from high variance and latency spikes, confirming that since flat consensus enforces dense synchronization, our hierarchy-aware orchestration is the essential prerequisite for enabling the physical buffer shrinkage that drives communication efficiency.

\subsubsection{Breakdown of Hierarchical Latency}
\begin{figure}[h]
    \centering
    \includegraphics[width=\linewidth]{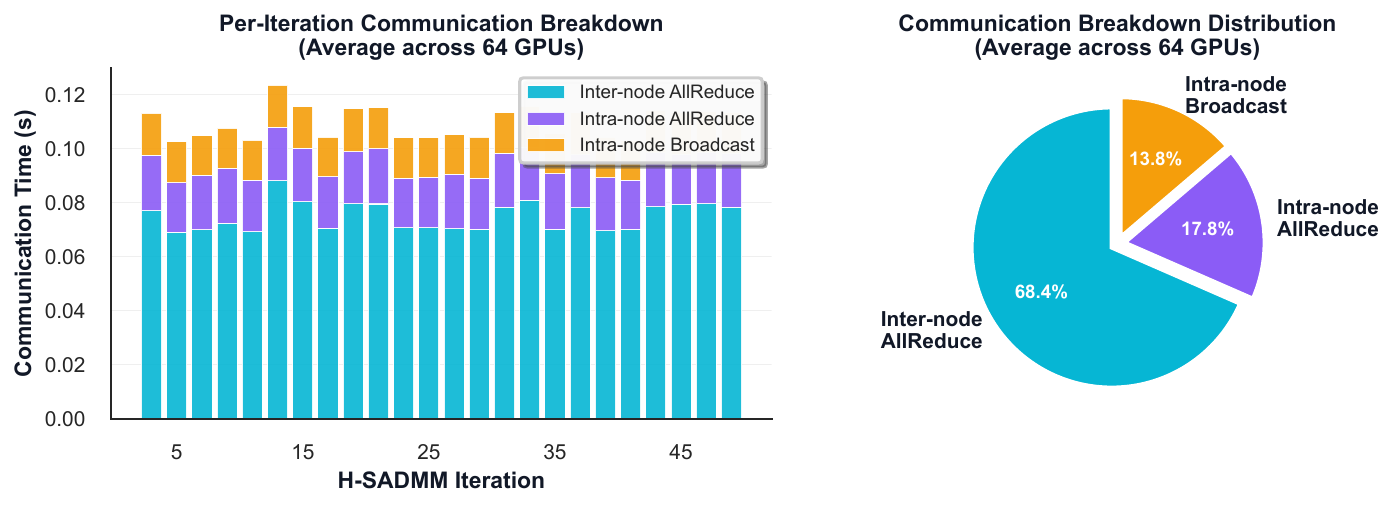}
    \caption{Breakdown of Communication Latency. Decomposition of communication time averaged across 64 GPUs, revealing that Inter-node \texttt{AllReduce} dominates the latency budget (68.4\%), validating the design choice to target inter-node traffic for compression.}
    \label{fig:hierarchical_breakdown}
\end{figure}

To verify the hierarchy-aware orchestration, we decompose the communication time in Figure~\ref{fig:hierarchical_breakdown}, averaged across all 64 GPUs. 
The breakdown reveals that the Inter-node \texttt{AllReduce} (cyan) dominates the latency budget, accounting for $68.4\%$ of the total communication time. 
The Intra-node \texttt{AllReduce} (purple) and \texttt{Broadcast} (orange) operations, which leverage high-bandwidth NVLink interconnects, consume $17.8\%$ and $13.8\%$ respectively.
This distribution empirically validates the design motivation behind \textsc{PruneX}. 
Since the inter-node step constitutes the primary system bottleneck due to slower physical interconnects, applying aggressive compression specifically to this stage allows \textsc{PruneX} to improve performance gains where they are most needed, rather than optimizing the already-fast local communication.

\subsection{Scalability Analysis}
\label{subsec:scalability}

\begin{figure}[t]
    \centering
    \includegraphics[width=\linewidth]{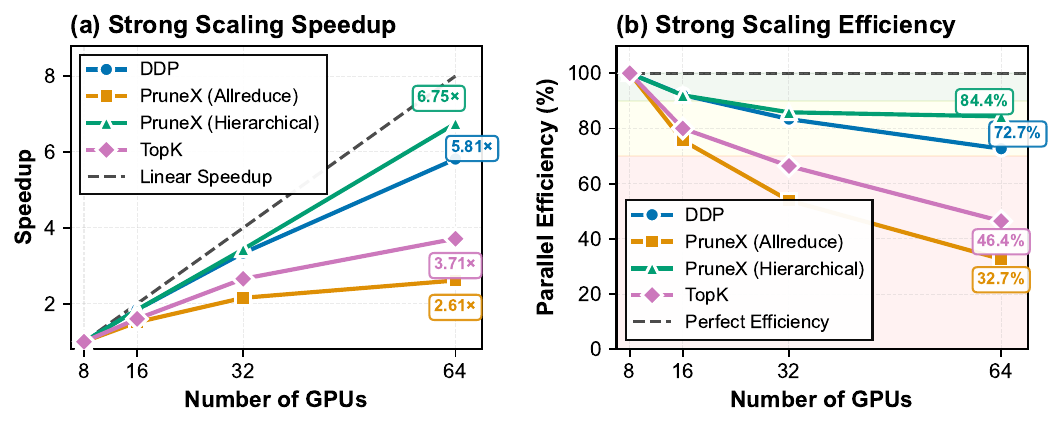}
    \caption{Strong Scaling Analysis. (a) Speedup and (b) Parallel Efficiency relative to an 8-GPU baseline, showing \textsc{PruneX} (Hierarchical) achieving 6.75x speedup on 64 GPUs and outperforming DDP and Top-K baselines. }
    \label{fig:scalability}
\end{figure}

To verify that the communication gains translate into actual system scaling, we conducted a strong scaling analysis by increasing the resource count from 8 GPUs (2 nodes) to 64 GPUs (16 nodes). 
Figure~\ref{fig:scalability} presents the speedup and parallel efficiency relative to the 8-GPU baseline. \textsc{PruneX} (Hierarchical) demonstrates superior scalability, achieving a $6.75\times$ speedup on 64 GPUs, corresponding to a parallel efficiency of $84.4\%$. In comparison, the standard dense DDP baseline achieves only $5.81\times$ speedup ($72.7\%$ efficiency), as the dense gradient synchronization becomes a bottleneck on the inter-node links.
Similarly, the Top-K baseline saturates at $3.71\times$ speedup, likely due to the computational overhead of sorting and the non-contiguous memory access patterns required for gradient sparsification.

\subsection{Algorithmic Convergence and Stability}
\label{subsec:eval_convergence}

A critical requirement for distributed ADMM is ensuring that the decoupled sub-problems actually converge to a consensus solution. We validate this by empirical analysis of the trajectory of the primal residuals, which measure the violation of the consensus constraints $\boldsymbol{\theta}_{i,j} = \mathbf{z}_i$ (intra-node) and $\mathbf{z}_i = \mathbf{z}$ (inter-node).

\subsubsection{Consensus Stability Across Ranks}
\begin{figure}[t]
    \centering
    \begin{subfigure}{\linewidth}
        \includegraphics[width=\linewidth]{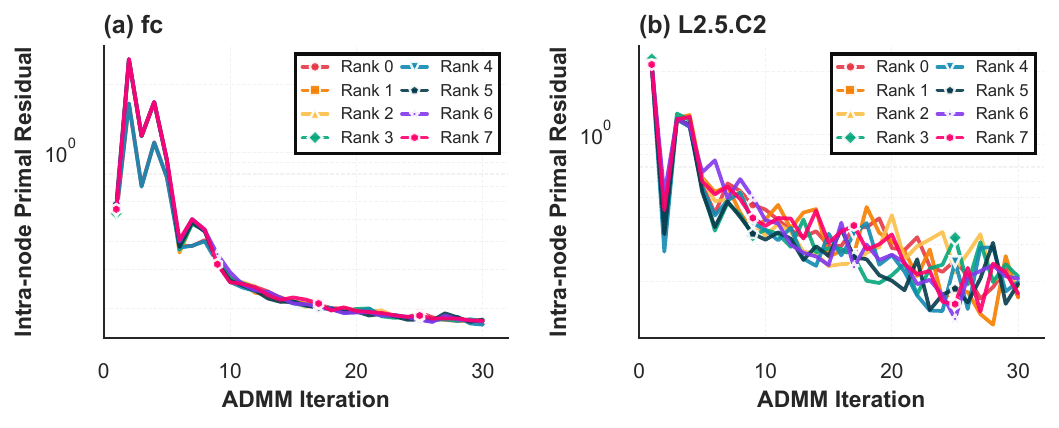}
        \caption{Intra-node Primal Residuals}
    \end{subfigure}
    \vfill
    \begin{subfigure}{\linewidth}
        \includegraphics[width=\linewidth]{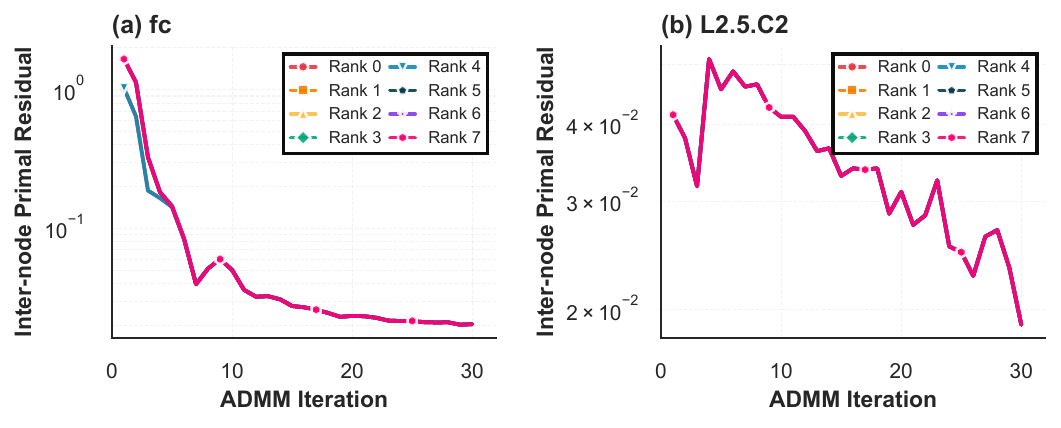}
        \caption{Inter-node Primal Residuals}
    \end{subfigure}
    \caption{Rank-wise Residual Convergence. Trajectories of (a) intra-node and (b) inter-node primal residuals for fully connected and convolutional layers. The consistent monotonic decay across all individual ranks  demonstrates consistent decay and successful consensus across all individual ranks.}
    \label{fig:rankwise_residuals}
\end{figure}

Figure~\ref{fig:rankwise_residuals} visualizes the primal residuals for individual ranks during the training of ResNet-152. Figure~\ref{fig:rankwise_residuals}(a) and (b) show the intra-node residuals $\|\mathbf{r}_{\text{intra}}\|_F$ for the fully connected (\texttt{fc}) and convolutional (\texttt{L2.5.C2}) layers, respectively. We observe a consistent monotonic decay across all ranks (0--7), driving the residuals from an initial magnitude of $10^0$ down to $10^{-2}$. This confirms that the local SGD updates on individual GPUs are successfully being pulled toward their respective node-level means by the penalty term $\rho_1^\ell$.
Similarly, Figure~\ref{fig:rankwise_residuals}(c) and (d) depict the inter-node residuals $\|\mathbf{r}_{\text{inter}}\|_F$. This indicates that the node leaders are effectively synchronizing the global state $\mathbf{z}$, and the system is converging to a valid global model where $\mathbf{z}_i \approx \mathbf{z}$ for all nodes.

\subsubsection{Layer-wise Convergence Dynamics}
\begin{figure}[t]
    \centering
    \includegraphics[width=\linewidth]{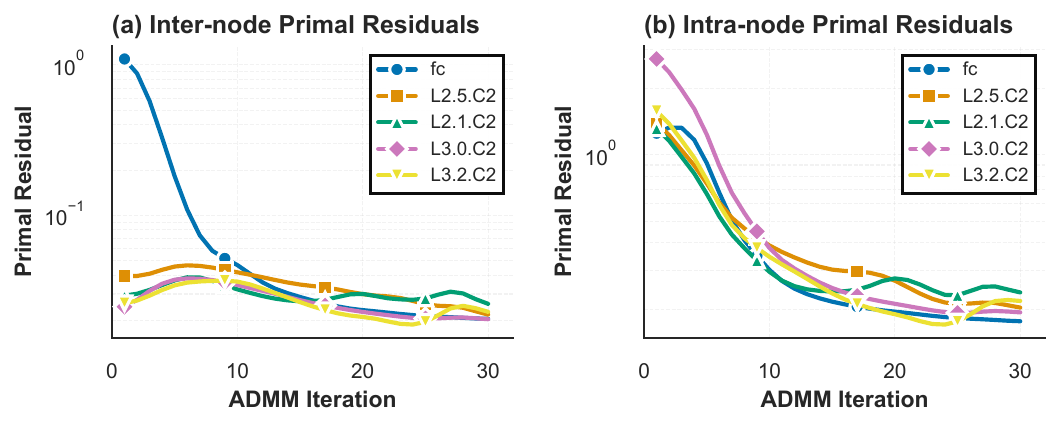}
    \caption{Layer-wise Primal Residuals. The distinct convergence rates of fully connected (blue) vs. convolutional layers (others) validate the necessity of the layer-wise adaptive penalty mechanism.}
    \label{fig:layerwise_residuals}
\end{figure}

Deep networks exhibit significant heterogeneity in parameter distributions across layers. Figure~\ref{fig:layerwise_residuals} plots the total primal residuals for the top-5 layers of ResNet-152. The fully connected layer (\texttt{fc}) exhibits a sharp initial drop, reflecting its dense connectivity and rapid adaptation. In contrast, the convolutional layers (e.g., \texttt{L2.5.C2}) show more gradual convergence profiles.

This disparity provides strong empirical justification for our \textit{layer-wise adaptive penalty tuning} strategy. A single global penalty $\rho$ would likely be too aggressive for the sensitive convolutional layers or too lax for the robust FC layers. By adapting $\rho^\ell$ independently, \textsc{PruneX} ensures that all layers converge synchronously, maintaining stability throughout the optimization process.

\subsection{Sparsity-Accuracy Trade-off}
\label{subsec:eval_accuracy}

Finally, we evaluate the quality of the compressed models produced by the H-SADMM algorithm. A central premise of \textsc{PruneX} is that deep networks possess significant redundancy, allowing for aggressive structured pruning without compromising representational power. 

\subsubsection{Pareto Frontier Analysis}
\begin{figure}[t]
    \centering
    \includegraphics[width=\linewidth]{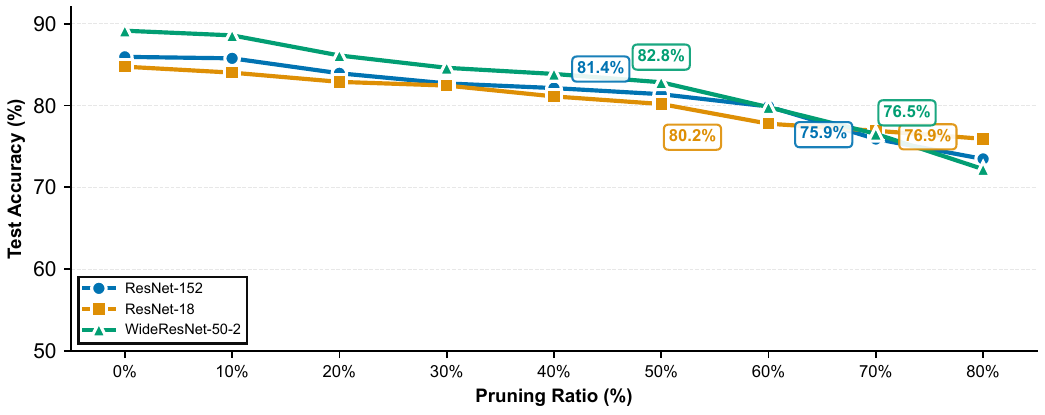}
    \caption{Sparsity-Accuracy trade-off. ResNet-152 maintains robust accuracy ($82.8\%$) even at $50\%$ structured (channel) sparsity. The swift drop-off only occurs in the hyper-sparse regime ($>70\%$), demonstrating the effectiveness of the H-SADMM projection in preserving critical features.
    }
    \label{fig:accuracy_sparsity}
\end{figure}

Figure~\ref{fig:accuracy_sparsity} depicts the Pareto frontier of test accuracy versus pruning ratio for three architectures on CIFAR-10. We observe a strong, but not uniform, resilience to channel pruning across all models. ResNet-152 retains a test accuracy of $81.39\%$ at $50\%$ pruning, corresponding to a drop of approximately $4.6$ percentage points from the unpruned baseline ($85.96\%$). Accuracy degradation becomes noticeable at moderate pruning levels: the drop exceeds $1\%$ by $20\%$ pruning and reaches about $3.8$ points at $40\%$ pruning. While pruning up to $50\%$ still preserves reasonably high performance, this regime cannot be considered strictly lossless. Importantly, the exact accuracy--sparsity trade-off depends on several factors, including hyperparameter tuning, optimizer choice, and the specific pruning criterion. Moreover, alternative strategies such as progressive or iterative pruning, longer fine-tuning, or combining pruning with complementary compression techniques could further mitigate accuracy loss and potentially improve performance at comparable sparsity levels. Similar qualitative trends are observed other models, though with different sensitivity profiles, confirming that the H-SADMM formulation exhibits consistent behavior across architectures.


\subsubsection{Visualization of Structured Support}
\begin{figure}[t]
    \centering
    \includegraphics[width=\linewidth]{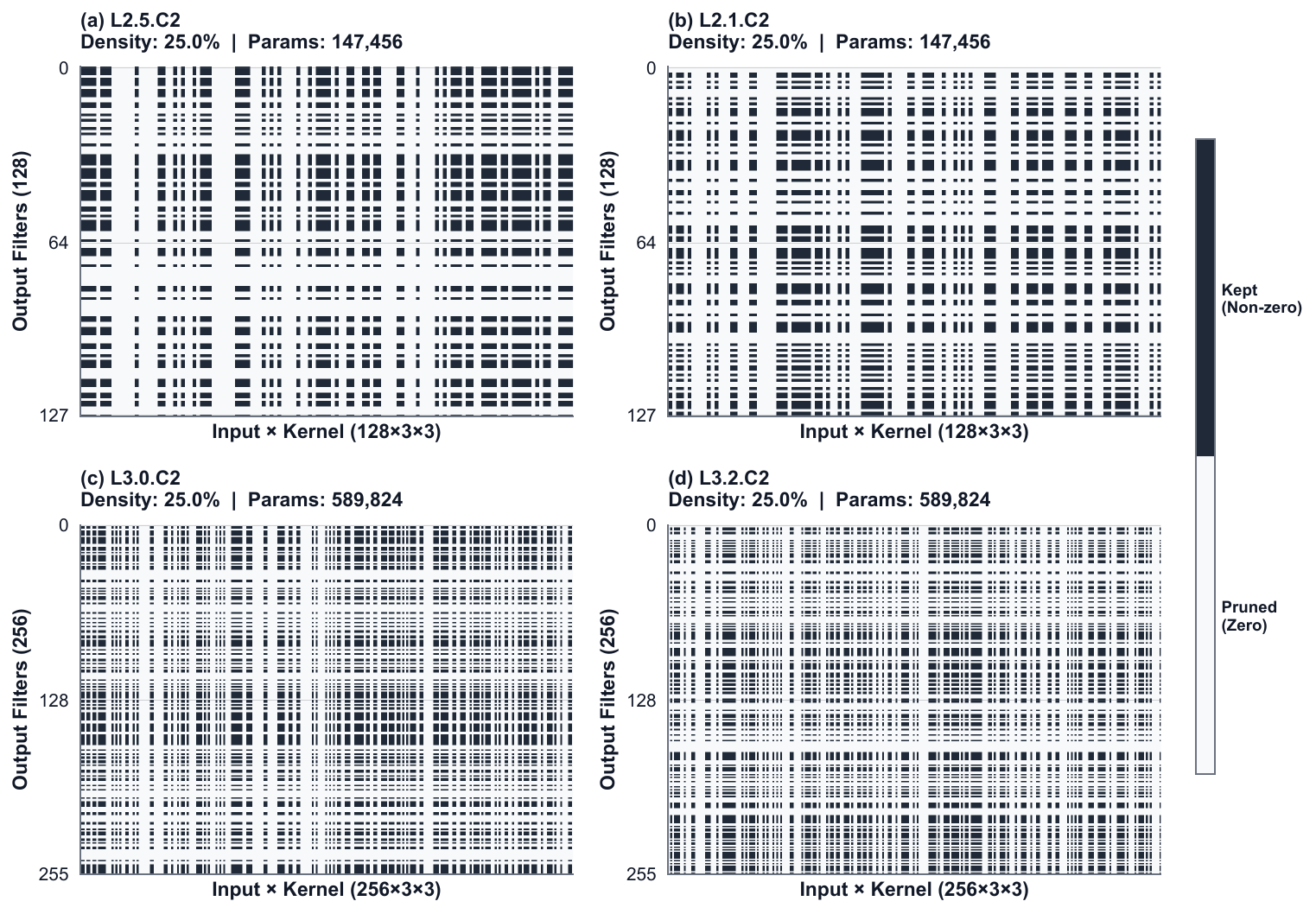}
    \caption{Visualization of Learned (Channel + Filter) Sparsity Patterns ($75\%$ Pruning). The white regions indicate pruned weights.
    }
    \label{fig:sparsity_patterns}
\end{figure}

To confirm that the imposed sparsity is indeed \textit{structured} Figure~\ref{fig:sparsity_patterns} visualizes the sparsity pattern for four representative convolutional layers of ResNet-152 at $75\%$ sparsity.
Unlike unstructured magnitude pruning, our masks exhibit distinct vertical and horizontal striations. A solid vertical white stripe corresponds to a completely pruned input channel ($\mathcal{S}_c$), while a horizontal stripe corresponds to a pruned output filter ($\mathcal{S}_f$). This block-wise structure is what enables \textsc{PruneX} to physically slice the tensors into smaller dense blocks ($\hat{\mathbf{c}}$) for efficient inter-node transmission, validating the end-to-end co-design of the optimization algorithm and the system architecture.

\section{Conclusion}
\label{sec:conclusion}
We presented \textsc{PruneX}, a hierarchy-aware distributed training system that couples structured pruning with hierarchical synchronization to reduce inter-node communication while preserving accuracy. By enforcing consistent global sparsity masks and physically shrinking inter-node communication buffers, \textsc{PruneX} translates model sparsity into concrete bandwidth savings and improved scalability. Across our evaluation, \textsc{PruneX} achieves reductions in inter-node traffic and communication latency compared to the baselines, while maintaining competitive convergence. Future work includes scaling \textsc{PruneX} along both the model and system axes. On the model side, we plan to extend the method beyond CNNs to larger architectures such as Vision Transformers (ViTs)~\cite{han2022survey,khan2022transformers} and other attention-based backbones~\cite{vaswani2017attention}. On the systems side, we aim to validate the approach on much larger GPU clusters (up to thousands of GPUs), where hierarchical communication becomes critical.


\section*{Acknowledgments}
This work was conducted within the Data Analytics for Zero Emission Marine (DAZE) project, with financial support from Business Finland (ref. 6525/31/2022). The authors also wish to acknowledge CSC – IT Center for Science, Finland, for providing computational resources.

\balance
\bibliographystyle{ACM-Reference-Format}
\bibliography{prunex}

\end{document}